\documentclass[twocolumn]{aastex631}

\usepackage{amsmath}
\usepackage{graphicx}
\usepackage{parskip}
\usepackage{float}
\usepackage{makecell}
\usepackage[utf8]{inputenc}
\usepackage{rotating}
\usepackage{mwe}

\newcommand{\be}{\begin{eqnarray}}
\newcommand{\ee}{\end{eqnarray}}

\newcommand{\src}{4U~1630--47}

\shorttitle{A Bayesian Framework for Spin Measurement}
\shortauthors{Das et al.}

\begin{document}
\title{Spin Constraints on 4U~1630--47 via combined Continuum Fitting and Reflection methods: a comparative study using Frequentist and Bayesian statistics}

\author[0009-0005-8097-7923]{Debtroy Das}
\affiliation{Center for Astronomy and Astrophysics, Department of Physics, Fudan University, Shanghai 200438, China}

\author[0000-0003-2845-1009]{Honghui Liu}
\altaffiliation{honghui.liu@uni-tuebingen.de}
\affiliation{Institut f\"ur Astronomie und Astrophysik, Eberhard-Karls Universit\"at T\"ubingen, D-72076 T\"ubingen, Germany}

\author[0000-0003-0847-1299]{Zuobin Zhang}
\affiliation{Astrophysics, Department of Physics, University of Oxford, Oxford OX1 3RH, UK}

\author[0000-0002-3180-9502]{Cosimo Bambi}
\altaffiliation{bambi@fudan.edu.cn}
\affiliation{Center for Astronomy and Astrophysics, Department of Physics, Fudan University, Shanghai 200438, China}
\affiliation{School of Humanities and Natural Sciences, New Uzbekistan University, Tashkent 100000, Uzbekistan}

\author[0000-0002-9639-4352]{Jiachen Jiang}
\affiliation{Department of Physics, University of Warwick, Coventry CV4 7AL, UK}

\author[0000-0003-0426-6634]{Johannes Buchner}
\affiliation{Max Planck Institute for Extraterrestrial Physics, D-85748 Garching, Germany}

\author[0000-0003-4187-9560]{Andrea Santangelo}
\affiliation{Institut f\"ur Astronomie und Astrophysik, Eberhard-Karls Universit\"at T\"ubingen, D-72076 T\"ubingen, Germany}
\affiliation{Center for Astronomy and Astrophysics, Department of Physics, Fudan University, Shanghai 200438, China}

\author[0000-0001-8250-3338]{Menglei Zhou}
\affiliation{Institut f\"ur Astronomie und Astrophysik, Eberhard-Karls Universit\"at T\"ubingen, D-72076 T\"ubingen, Germany}

\begin{abstract}
We present a comprehensive Bayesian spectral analysis of the black hole X-ray binary 4U~1630–47 during its 2022 outburst, using simultaneous \textit{NICER} and \textit{NuSTAR} observations. Using the traditional frequentist approach, we build our model combining reflection spectroscopy with continuum fitting techniques and analyze the data. In the Bayesian framework, we jointly constrain the black hole's spin, mass, and inclination within a unified framework. Employing nested sampling, we capture parameter degeneracies and rigorously propagate both statistical and systematic uncertainties. Our results yield robust evidence for a high spin, with the exact value still subject to model-dependent systematic uncertainties from both approaches. Our Bayesian analysis infers spin $a^* = 0.96_{-0.03}^{+0.02}$, mass $M_{\rm BH} = 12.19_{-1.46}^{+1.22} \, M_\odot$, and inclination angle $i = 55.75_{-1.37}^{+1.60}$~degrees, obtained via \texttt{relxillNS} flavor of our model. It also demonstrates the power of Bayesian inference in fetching valuable insights into the complex physics of black hole accretion and enabling high-confidence measurements of fundamental parameters.
\end{abstract}


\section{Introduction}
\label{intro}
Astrophysical black holes (BHs) are completely described by two parameters: the mass and the spin. 
Stellar-mass (3-150 solar masses, $M_{\odot}$) BHs are often found in binary systems \citep{Bambi:2025rod}, and supermassive BHs ($>10^5$~$M_{\odot}$) are found to reside in the center of almost all galaxies. Measuring spins of stellar-mass BHs is important for understanding their formation mechanisms \citep{Reynolds2021ARA&A..59..117R}. The spins of supermassive BHs are thought to be related to their merger and accretion histories \citep{Volonteri2005ApJ...620...69V, Berti2008ApJ...684..822B}. The spins of stellar-mass BHs may reflect the physics of core-collapse of massive stars \citep{1999MNRAS.305..654K,2010Natur.468...77V,2012ApJ...747..111W,Miller2015PhR...548....1M}; but see also \citep{Fragos:2014cva,2019ApJ...870L..18Q}. Moreover, BH spins are believed to govern some important high-energy astrophysical phenomena such as the relativistic jets \citep{Penrose1969NCimR...1..252P, Blandford1977MNRAS.179..433B}.

For BH X-ray binary (XRBs) systems, X-ray reflection spectroscopy \citep{Brenneman2006ApJ...652.1028B, Reynolds2014SSRv..183..277R, Bambi2021} and continuum fitting \citep{Zhang1997ApJ...482L.155Z, McClintock2014SSRv..183..295M} are the two leading methods to measure BH spins. These methods rely on modeling the thermal and reflection components of the X-ray spectrum. The accretion disk emits radiation that is locally blackbody-like, resulting in a multi-temperature blackbody spectrum as a whole \citep{Shakura1973, Mitsuda1984}. The temperature profile of the disk depends on the BH spin, mass, and mass accretion rate \citep{Novikov1973blho.conf..343N}. A hot corona can Compton up-scatter seed photons from the disk or internally the corona itself, producing a non-thermal power-law-like spectrum with a high-energy cutoff \citep{Shapiro1976ApJ...204..187S, Galeev1979ApJ...229..318G}. A fraction of the coronal photons could be reprocessed by the accretion disk, leading to the reflection component. The local reflection spectrum is characterized by fluorescent emission lines, the most prominent one is normally the iron K$\alpha$ line at 6.4--7.0 keV, and by a Compton hump peaked around 30 keV \citep{George1991, Garcia2010}. The lines and the continuum are then blurred by relativistic effects near the black hole (Doppler effect and gravitational redshift), causing a prominent broad line feature in the iron band \citep[e.g.,][]{Fabian1989, Walton2016, Jiang2019gx339, Liu2022MNRAS.513.4308L, Liu2023ApJ...950....5L}. Information about the BH spin and the accretion flow is encoded in the broadening effects \citep[e.g.,][]{Fabian2000, Dauser2010, Liu2019, Bambi2021, Liu2023ApJ...951..145L}.

Another potential method to measure BH spins in XRBs is to associate the frequencies of quasi-periodic oscillations (QPOs, \citealt{Ingram2019NewAR..8501524I}) with dynamical frequencies of the accretion system. For instance, in the relativistic precession model (RPM), the frequencies of Type-C QPO and the lower and upper high-frequency QPOs are thought to correspond to the Lense-Thirring precession, the periastron precession, and the orbital frequencies at a certain radius for a test particle \citep{Motta2014MNRAS.437.2554M, Motta2014MNRAS.439L..65M, Bhargava2021MNRAS.508.3104B, Motta2022MNRAS.517.1469M, Motta2024A&A...684A.209M}. These frequencies are determined by the BH mass and spin parameters \citep{Stella1998ApJ...492L..59S, Stella1999PhRvL..82...17S, Stella1999ApJ...524L..63S}. Simultaneous detections of all three QPOs (or two of them) are rare and have only been found in a handful of sources where the RPM can be applied to constrain BH spins (see Table 1 of \citealt{Mall2024MNRAS.52712053M} and references there in). It should be noted that the physical origin of QPOs remains a topic of debate \citep{Ingram2019NewAR..8501524I}. Even if QPOs originate from the precession of the accretion flow, purely test-particle dynamics may not provide an accurate description of the system \citep{Fragile2007ApJ...668..417F,Ingram2009MNRAS.397L.101I, Motta2018MNRAS.473..431M}. Moreover, alternative QPO models that do not involve precession have also been developed \citep[e.g.,][]{Tagger1999A&A...349.1003T}, and some of them can also be used to measure BH spins \citep{Dhaka:2023vxv,Misra2020ApJ...889L..36M, Liu2021ApJ...909...63L, Zhang2024ApJ...971..148Z}. As of now, we do not know which model is the correct one, and BH spin measurements from QPOs usually disagree with those inferred from X-ray reflection spectroscopy and continuum fitting \citep[see, for example, the discussion in][]{Mall2024MNRAS.52712053M}. 

The reflection and continuum fitting methods are often used separately to constrain BH spins, although measurements from both methods indicate relatively high BH spins for BH XRBs \citep[e.g.,][]{Draghis2023ApJ...946...19D, Draghis2024ApJ...969...40D}. This differs from the results of gravitational wave sources, where low BH spins are preferentially found \citep{Fishbach2022ApJ...929L..26F, Abbott2023PhRvX..13a1048A}. The discrepancy may indicate two different populations of stellar-mass BHs \citep{2022ApJ...929L..26F}, but it is also possible that one of the two methods (or both) does not provide an accurate measurement of the spin distribution. Indeed, the X-ray methods of measuring BH spins could be impacted by our understanding of the complex accretion physics and limitations in spectral model construction \citep[see][for a recent review]{Zdziarski2025arXiv250600623Z}. 

Using the continuum fitting method alone to constrain BH spins requires prior knowledge of the BH mass, distance, and the disk inclination angle. This limits the number of sources for which spin can be measured with this method. Combining continuum fitting with the reflection method could help relax these prior requirements, because the reflection method can bring extra constraints on the disk inclination angle and BH spin. In this way, it is also possible to estimate the BH mass using only X-ray data \citep[e.g.,][]{Parker2016ApJ...821L...6P,Tripathi:2020yts,Zhang2022ApJ...924...72Z, Zdziarski2025ApJ...981L..15Z}. Such an analysis requires data in the appropriate spectral state, where the thermal emission is strong enough for continuum fitting to be applied and where reflection features are also present.


\begin{figure*}
	\centering
	
		\includegraphics[width=\linewidth]{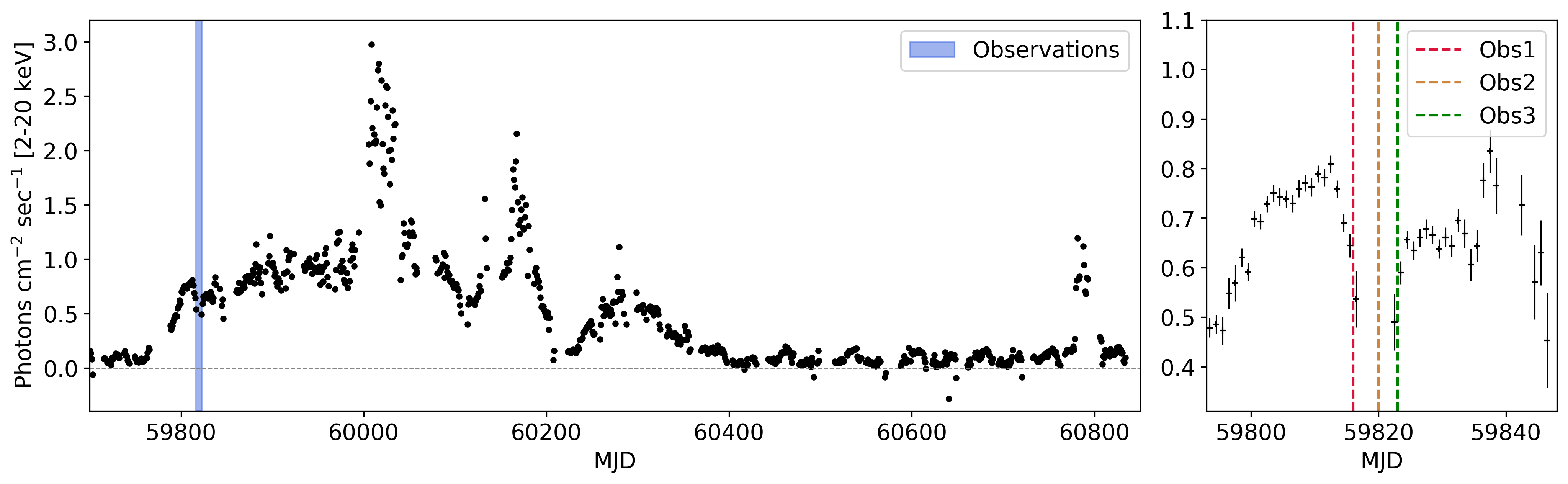} 
	
	\caption{(Left Panel) 1-day-averaged MAXI light curves of \src{} from May 2022 to May 2025 showing several outbursts. The 2022 X-ray outburst is indicated by the blue-shaded region. (Right Panel) Zoomed-in version of the 1-day-averaged MAXI light curve around the time of the 2022 outburst. Our observation dates are marked by the red, yellow, and green vertical dashed lines. Their MJDs are 59816, 59820, and 59823, respectively, which correspond to 2022-08-25, 2022-08-29, and 2022-09-01.}
	\label{fig:lc}
\end{figure*}

In this paper, we apply the continuum fitting and reflection methods to \src{}, using \textit{NICER} and \textit{NuSTAR} observations taken during its soft state in the 2022 outburst. \src{} is a transient BH XRB system discovered in 1969. It exhibits frequent outbursts with a typical interval of 600 to 690 days \citep{Abe:2005nw,Tomsick:2014yva}. Since the black hole mass and disk inclination angle are unknown, previous spin measurements have been based on the reflection method, and a high spin was generally found \citep{Liu:2021tyw, King:2013fma}. Note that the source has been observed by the Imaging X-ray Polarimetry Explorer (\textit{IXPE}, \citealt{Soffitta2021AJ....162..208S, Weisskopf2022JATIS...8b6002W}) in 2022 in the soft and steep power-law states, and an unexpectedly high polarization degree was found \citep{Kushwaha2023MNRAS.524L..15K, Ratheesh2024ApJ...964...77R}.

In addition to the standard $\chi^2$ minimization, we also employ a Bayesian framework for parameter estimation and model comparison. Through the computation of Bayesian evidence, the Bayesian framework enables a rigorous comparison between competing models \citep{kass1995bayes}, allowing us to quantify the relative support for different physical scenarios. In addition, incorporating prior information within a Bayesian framework offers several advantages for parameter estimation. It ensures that the inferred parameters remain consistent with our existing knowledge of the system and helps to break the degeneracies that can arise between strongly correlated parameters \citep{gelman1995bayesian}, such as spin, inclination, and mass. Most previous X-ray spectral modeling efforts have not systematically incorporated such priors, even when reliable external constraints were available. Hence, our approach provides a more coherent and physically informed estimation of the BH parameters.

The following section describes our observational data and the data reduction processes. Section 3 presents a detailed methodology used in the reflection spectroscopy and continuum fitting methods, along with the best-fit estimates of the parameters. We portray the Bayesian analysis in Section 4 by discussing the mathematical foundation and our strategy for the analysis. We also demonstrate the results by providing posterior distributions for the BH spin and other system parameters before discussing how our findings compare with previous studies. Finally, we offer conclusions and potential directions for future research, highlighting the significance of our approach for understanding BH systems such as 4U~1630-47. 

\section{Observation and data reduction}\label{obs}

\begin{figure} [h]
    \centering
    \includegraphics[width=\linewidth]{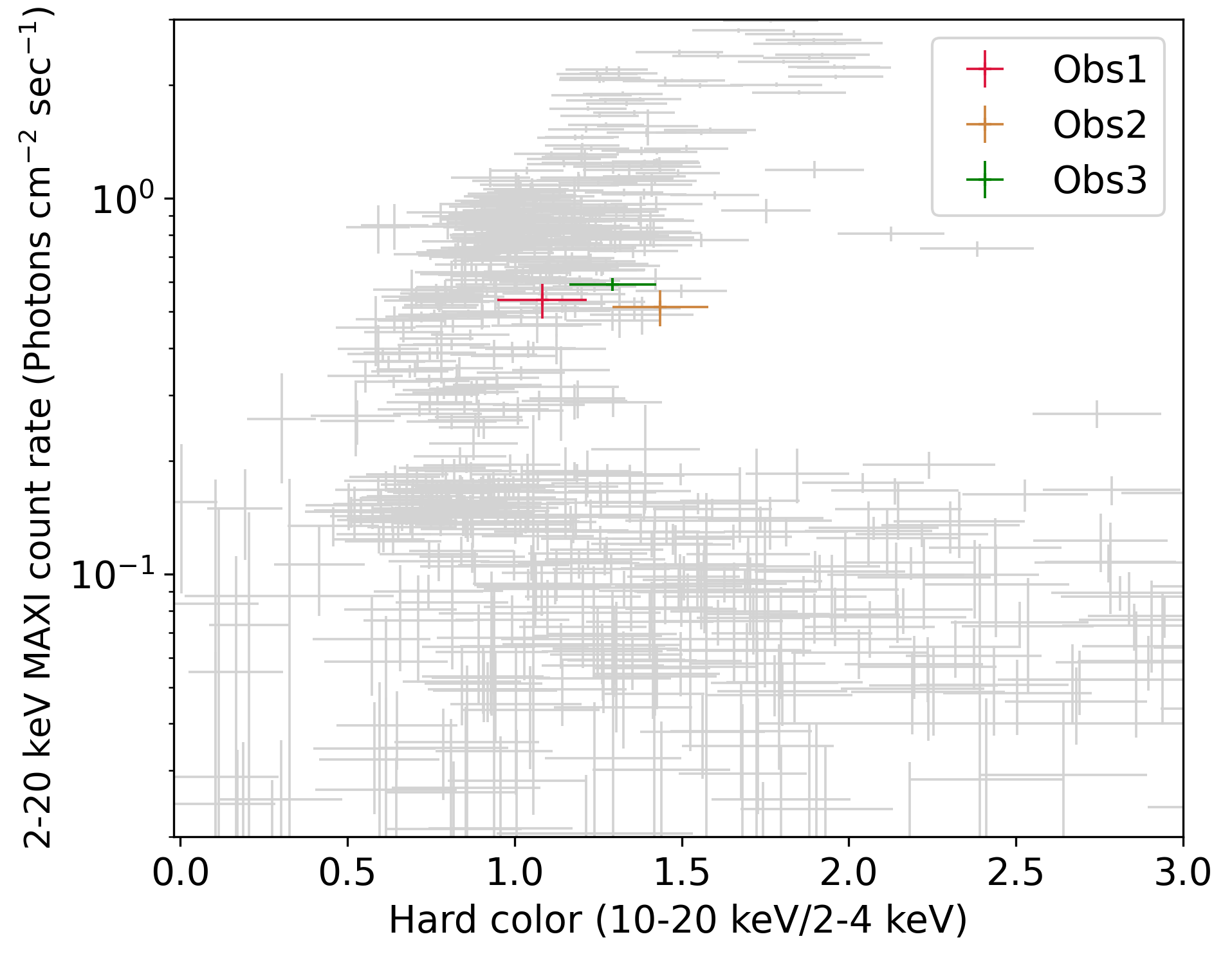}
    \caption{The hardness-intensity diagram using 1-day-averaged MAXI light curves of \src{}. The three observations considered in this work are marked with colors.}
    \label{fig:hid}
\end{figure}

\src{} went into an outburst in 2022 (see Figure~\ref{fig:lc}). We chose three \textit{NuSTAR} observations during its rising phase in the spectrally soft state. The state can be distinguished through the hardness-intensity diagram in Figure~\ref{fig:hid}. For each \textit{NuSTAR} observation, we also include a simultaneous \textit{NICER} observation to cover the soft energy band. Details of the observations considered are shown in Table~\ref{obslog}.

\begin{table} [h]
    \centering
    \caption{Log of \textit{NICER} and \textit{NuSTAR} observations of 4U~1630-47 considered in the analysis. From left to right, (1) name of observation; (2) observation ID; (3) start MJD of observation; (4) exposure time}
    \label{obslog}
    \renewcommand\arraystretch{1.5}
    \begin{tabular}{llll}
        \hline\hline
        Obs Name & ObsID & Time & Exposure\\
        &&&(ksec) \\
        \hline
        &&\textit{NICER}&\\
        \hline
        ni1 & 5501010104 & 2022-08-25 01:42:36 & 3.9\\
        ni2 & 5501010108 & 2022-08-29 06:43:00 & 3.0 \\
        ni3 & 5501010111 & 2022-09-01 01:27:00 & 2.6 \\
        \hline
        &&\textit{NuSTAR}&\\
        \hline
        nu1 & 80802313002 & 2022-08-25 04:31:09 & 16.3\\
        nu2 & 80802313004 & 2022-08-29 08:16:13 & 13.2\\
        nu3 & 80802313006 & 2022-09-01 05:31:12 & 14.8\\
        \hline\hline
    \end{tabular}
    \\
    \textit{Note.} Obs 1 consists of ni1 and nu1; Obs 2 consists of ni2 and nu2; Obs 3 consists of ni3 and nu3. 
\end{table}

\subsection{\textit{NICER}}

The \texttt{nicerl2} task is first applied to process the \textit{NICER} data using the CALDB v20221001. Then we use the \texttt{nicerl3-spect} tool to extract source spectra. The standard default value used in the CALDB systematic error vector for \texttt{nicerl3-spect} is 1.0\%. The default background model \texttt{SCORPEON} is used to generate the background spectral file. The spectra are re-binned according to the optimal binning scheme \citep{kaastra2016optimal}, and each energy bin contains at least 30 counts. We have also tested an alternative model \texttt{3C50}, introduced a small fractional systematic uncertainty of 2\% to the spectra, and found that it has no impact on the black hole spin and disk inclination angle measurements.

\subsection{\textit{NuSTAR}}

The \textit{NuSTAR} data is processed with NuSTARDAS v2.1.1 and using \textit{NuSTAR} CALDB v20230307.
We use the \texttt{nupipeline} tool to extract cleaned event files for both FPMA and FPMB. The source spectra are extracted from a circular region centered on the source with a radius of 150\arcsec. The background spectra are extracted from a source-free circular region with a radius of 180\arcsec. We use the \texttt{nuproducts} to generate the spectra and other products. The FPMA and FPMB spectra are grouped to have a minimum count of 30 photons per bin.

\section{Spectral analysis with $\chi^2$ statistics}\label{s-ana}

Spectral fittings are conducted with \texttt{XSPEC} v12.13.0c \citep{xspec}.  $\chi^2$ statistics is used to find the best-fit values and uncertainties of the model parameters (at the $90\%$ confidence level unless otherwise specified). 

\subsection{Model build up}

We provide a detailed description of each component of the model used to fit the data and discuss the strategy behind its sequential inclusion. We start with Obs 1 and fit a phenomenological model: \texttt{TBabs*kerrbb}. The \texttt{TBabs} \citep{Wilms:2000ez} model is used to describe the absorption by the interstellar medium along the line of sight. We model the thermal emission from the accretion disk using \texttt{kerrbb}, which describes the multicolor blackbody emission from a thin accretion disk around a Kerr black hole \citep{Li:2004aq}. The torque at the disk's inner boundary is fixed at 0. We fix the black hole mass at $M_{\rm BH} = 10~\mathrm{M}_\odot$, and the distance at $d_{\rm BH} = 10$~kpc, based on previous studies by \citet{Kuulkers:1997fh,Seifina:2014ura,Pahari:2018toe,Kushwaha2023MNRAS.524L..15K}. We set \texttt{rflag=1} to enable the effect of self-irradiation (without using the \texttt{bhspec} model) and \texttt{lflag=1} to assume disk emission is limb-darkened. The other free parameters of \texttt{kerrbb} are the black hole spin ($a_*$), the disk inclination angle ($i$), and the mass accretion rate ($\dot{M}_{\rm BH}$).

\begin{figure}[h]
    \centering
    \includegraphics[width=\linewidth]{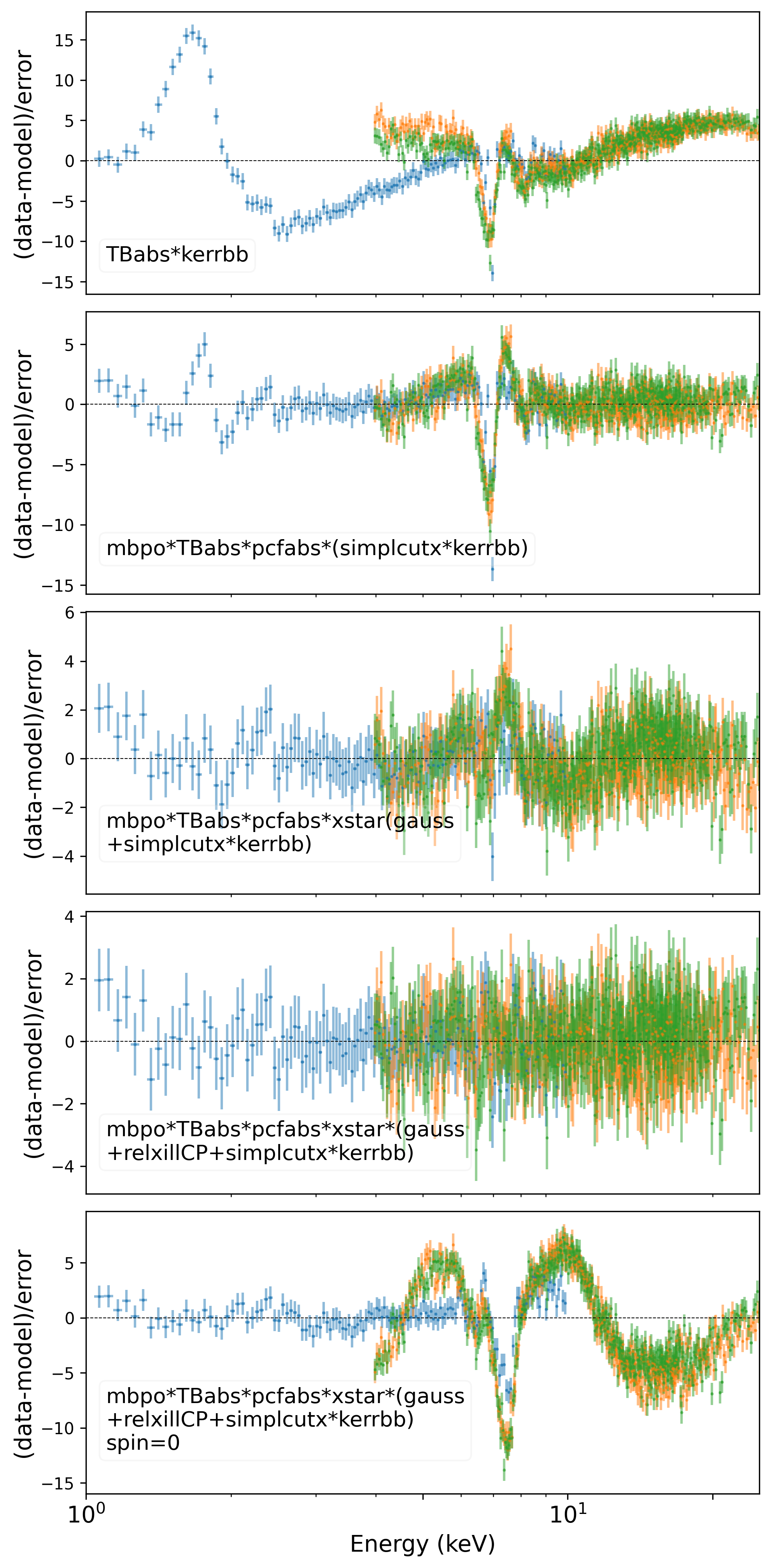}
    \caption{The residuals (in units of $\sigma$) for several models applied to the \textit{NICER} (in blue) and \textit{NuSTAR} (FPM-A in orange and FPM-B in green) spectra of \src{} obtained from Obs 1. The corresponding models are mentioned in each panel. In the bottom panel, we have forcefully set spin $a_*=0$ to demonstrate the impact of spin on the fit.}
    \label{model}
\end{figure}

The plot of the residuals of the fit (the first panel of Figure.~\ref{model}) shows several unresolved features. We include the \texttt{simplcutx} model to account for a possible Comptonization component \citep{Steiner2017}. We fix the reflection fraction parameter of \texttt{simplcutx} to $1$ and leave the photon index ($\Gamma$), the coronal temperature ($kT_{\rm e}$), and the scattering fraction ($f_{\rm sc}$) as free parameters. $kT_{\rm e}$, when left free, remains unconstrained, so we set the value to 100~keV. The model \texttt{mbpo} \citep{Ingram2017MNRAS.464.2979I} is used to cross-calibrate data from different instruments. It multiplies the total model by a constant (Norm), and a broken power-law with the index $\Delta\Gamma_{1}$ for $E<E_{\rm br}$ and $\Delta\Gamma_{2}$ for $E>E_{\rm br}$, where $E_{\rm br}$ is the breaking energy. For the \textit{NuSTAR} spectra, we fix the power-law indices to zero. The normalization is set to unity for the FPMA spectrum, but kept free for the FPMB spectrum.  For the \textit{NICER} data, we leave all \texttt{mbpo} parameters free. A prominent feature at low energy still exists. A partial covering component \texttt{pcfabs} is then included in the model, as motivated by \citep{Kushwaha2023MNRAS.524L..15K}.

Moreover, the gold emission line from the X-ray optics of \textit{NICER} is corrected with the inclusion of a narrow Gaussian component around 1.7 keV using \texttt{gauss} model \citep{Kushwaha2023MNRAS.524L..15K}. Additional absorption features are clearly visible in the residuals between 6 and 7 keV. These disk-wind-originating absorption lines are previously reported and attributed to rest frame energies of the Fe XXV and Fe XXVI lines at 6.697 keV and 6.966 keV, respectively \citep{Bianchi:2004eh,King:2013fma,Pahari:2018toe}. We use a custom photoionization table generated by the \texttt{xstar} code \citep{Kallman2001ApJS..133..221K} to model these absorption features. We consider that the illumination of the photoionized absorber is due to the absorption-corrected full continuum. For each observation, we extract the best-fit continuum model from the preceding fit. Given that there is only limited variation in flux and spectral shape across observations, we average the best-fit continuum models for each observation to obtain an averaged continuum model. It is used as the ionizing continuum for the photoionized absorber. The turbulent velocity of the plasma is set to be 1000~km~s$^{-1}$ and the density to 10$^{12}$~cm$^{-3}$. We developed custom photoionization tables generated by the \texttt{xstar} code with fixed turbulent velocities of 250, 500, 1000, 2000, and 4000~km~s$^{-1}$. When these tables were used one by one in our model, we found that the effect of $v_{turb}$ is minor on the estimates of the other system parameters.

In this step, we observed reflection features in the fit residuals. The residuals show structures near the iron line region and the Compton hump, indicating the need to add a reflection model to properly account for these physical processes. Hence, we used \texttt{relxillCp} to account for them \citep{Garcia:2013lxa}. $R_{\rm in}$ is fixed at the ISCO radius. $R_{\rm out}$ is fixed at 400 $r_{\rm g}$. The emissivity index, when left as a free parameter, remains unconstrained, so we set it to 3 for our parameter estimation.

Our final model for analysis takes the form: \texttt{mbpo * TBabs * pcfabs * xstar * (gauss + relxillCp + simplcutx * kerrbb)}. The black hole spin and disk inclination angle parameters of the \texttt{relxillCp} model are linked to those in the \texttt{kerrbb} model. The parameters for the incident coronal emission ($\Gamma$ and $kT_{\rm e}$) are linked to those in the \texttt{simplcutx} model. The reflection fraction parameter in \texttt{relxillCp} is set to $-1$. Thus, the model only returns the reflection component. The other parameters are the disk ionization parameter ($\log(\xi)$), the iron abundance in units of the solar value ($A_{\rm Fe}$), the disk electron density ($n_{\rm e}$), and the normalization parameter. This model combines the constraining power of both the continuum fitting and the reflection methods. Similar models have been applied to MAXI J1820+070 to understand the accretion geometry \citep{Fan2024ApJ...969...61F}, and to GRS 1716--249 to test general relativity \citep{Zhang2022ApJ...924...72Z}.

This model provides a good fit to the data, with $\chi^2/\nu= 944.6/909$, and there are no noticeable residuals (second from bottom panel of Figure~\ref{model}). Then we apply the final model to all three observations. We obtain a good fit for each observation. All the best-fit parameter values are listed in Table~\ref{CP}. The residuals to the best-fit models are shown in the left panels of Figure~\ref{fig:delchi}. With this model, we measure a high BH spin ($a_*>0.93$) and a relatively high disk inclination angle ($i\sim 50^{\circ}$). The power-law emission is weak, with a scattering fraction $<$ 5\% and a very soft photon index ($\Gamma>2.8$). We also obtain a disk density around 10$^{18}$~cm$^{-3}$.

The continuum-fitting method usually requires the black hole distance ($d_{\rm BH}$), inclination ($i$), and mass ($M_{\rm BH}$) to be fixed in order to constrain the spin $a_*$ and mass accretion rate $\dot{M}_{\rm BH}$ \citep[e.g.,][]{Gou2011ApJ...742...85G}. In our joint continuum-fitting and reflection analysis, the disk inclination and spin can be independently constrained by the reflection component. Therefore, we can allow additional degrees of freedom in the \texttt{kerrbb} model. However, due to the intrinsic degeneracy among $M_{\rm BH}$, $d_{\rm BH}$, and $\dot{M}_{\rm BH}$ in the continuum-fitting method, the X-ray data alone cannot constrain all three parameters simultaneously (see Figure 8 of \citealt{Parker:2019ryb}). Therefore, one of these three parameters should be fixed based on an independent measurement. Since the distance to 4U~1630--47 is known to be $\sim 10$~kpc based on modeling of its dust-scattering halo \citep{Kalemci:2025psu}, we fix the source distance and allow only $M_{\rm BH}$ and $\dot{M}_{\rm BH}$ to vary. The best-fit parameters and uncertainties are shown in Table~\ref{CP}, yielding a black hole mass of $6.5$--$10.5~M_{\odot}$.

\subsection{Impact of reflection models}
\label{subsec:impact}

\begin{figure*}[h]
	\centering
	\includegraphics[width=0.49\textwidth]{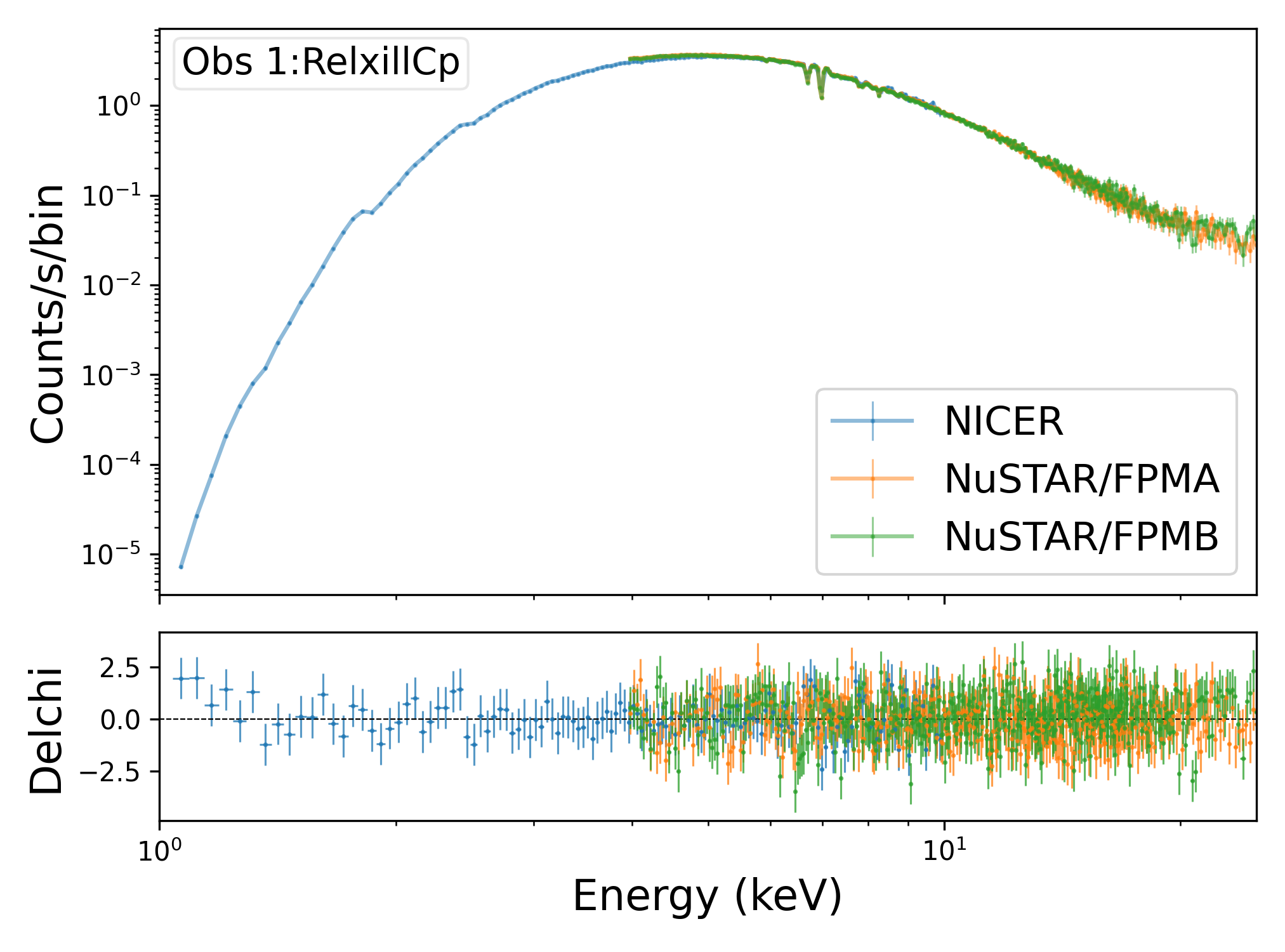}
	\centering
	\includegraphics[width=0.49\textwidth]{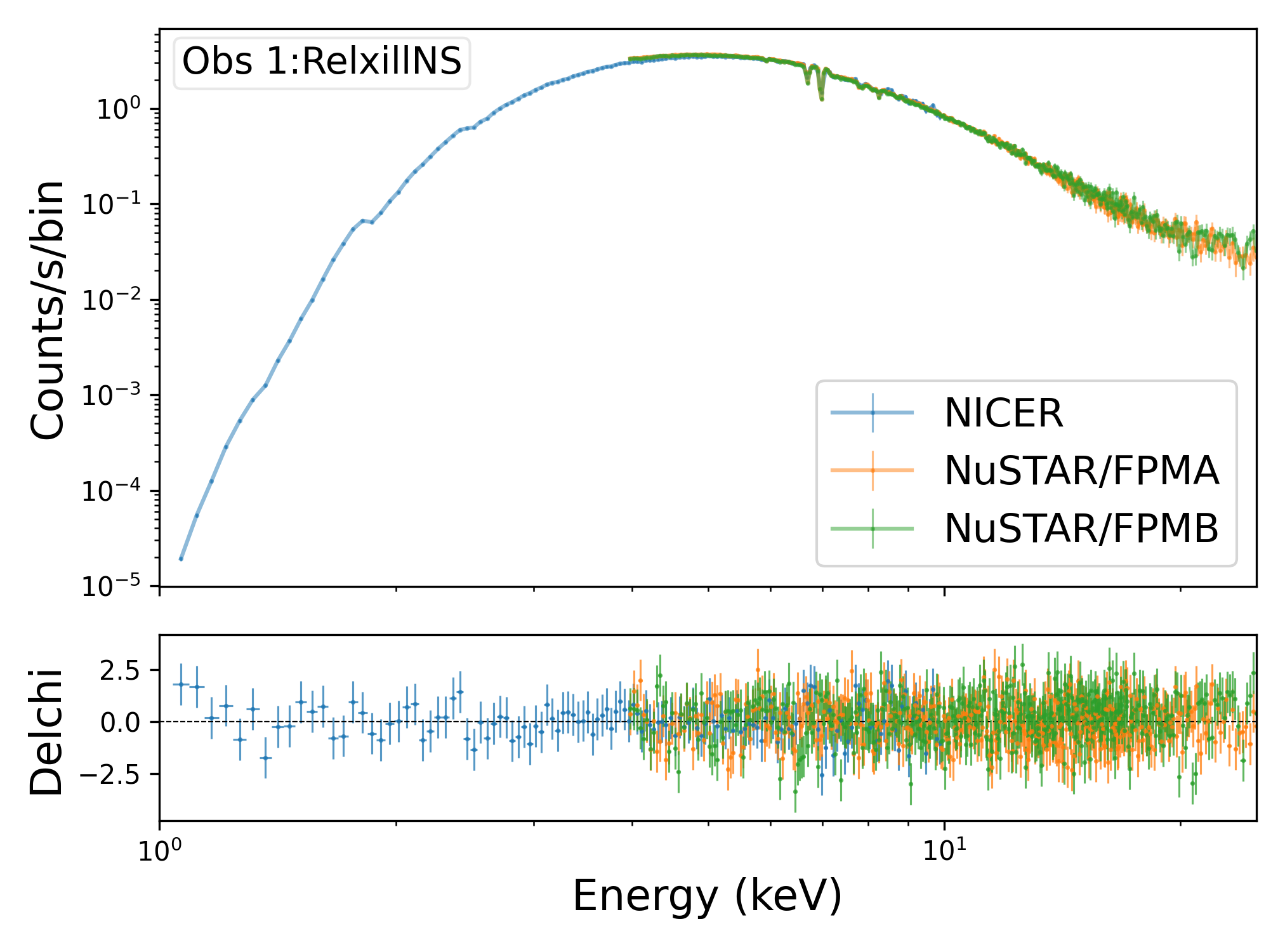}
	\vskip\baselineskip
	\centering
	\includegraphics[width=0.49\textwidth]{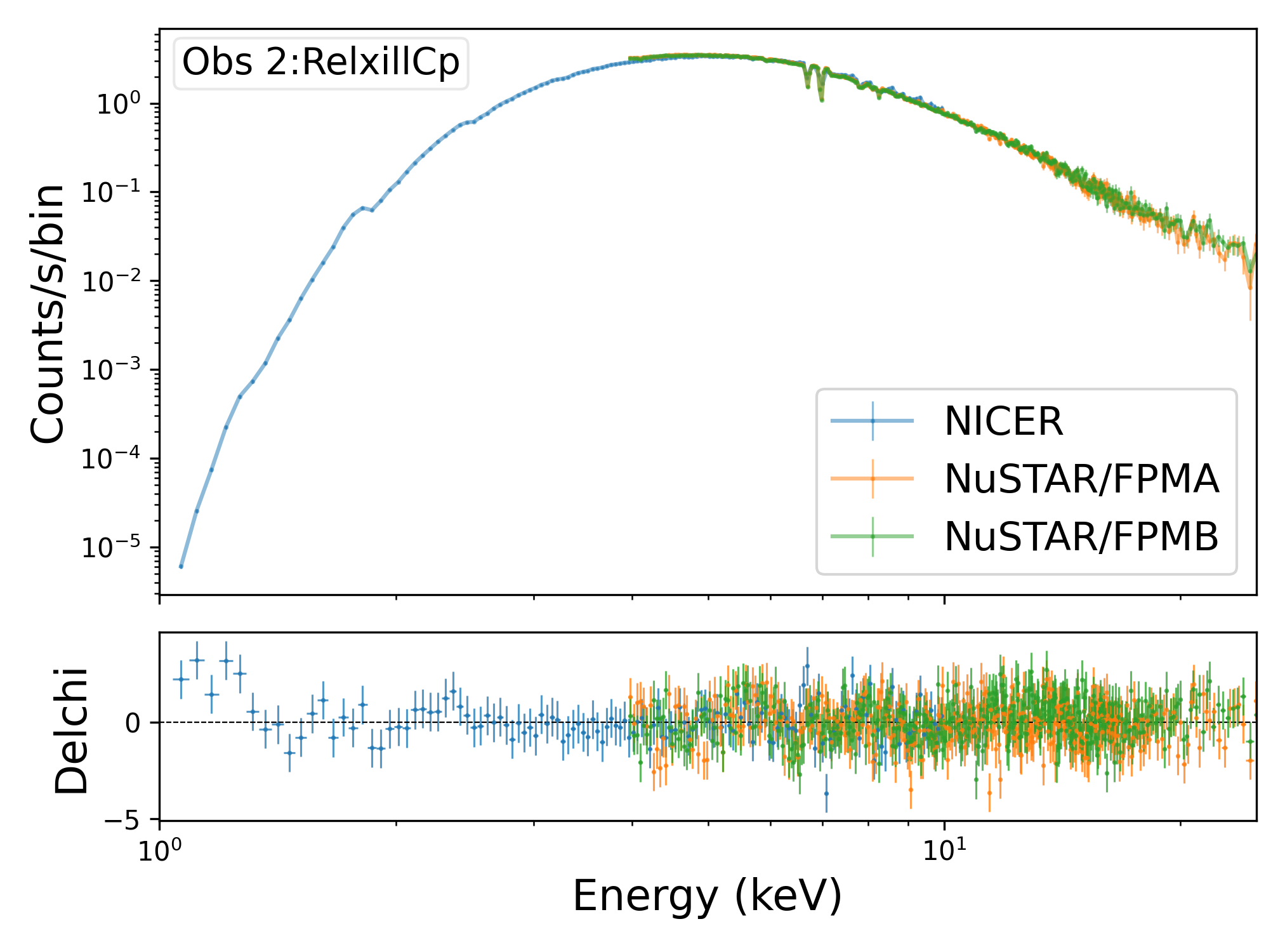}
	\hfill
	\centering
	\includegraphics[width=0.49\textwidth]{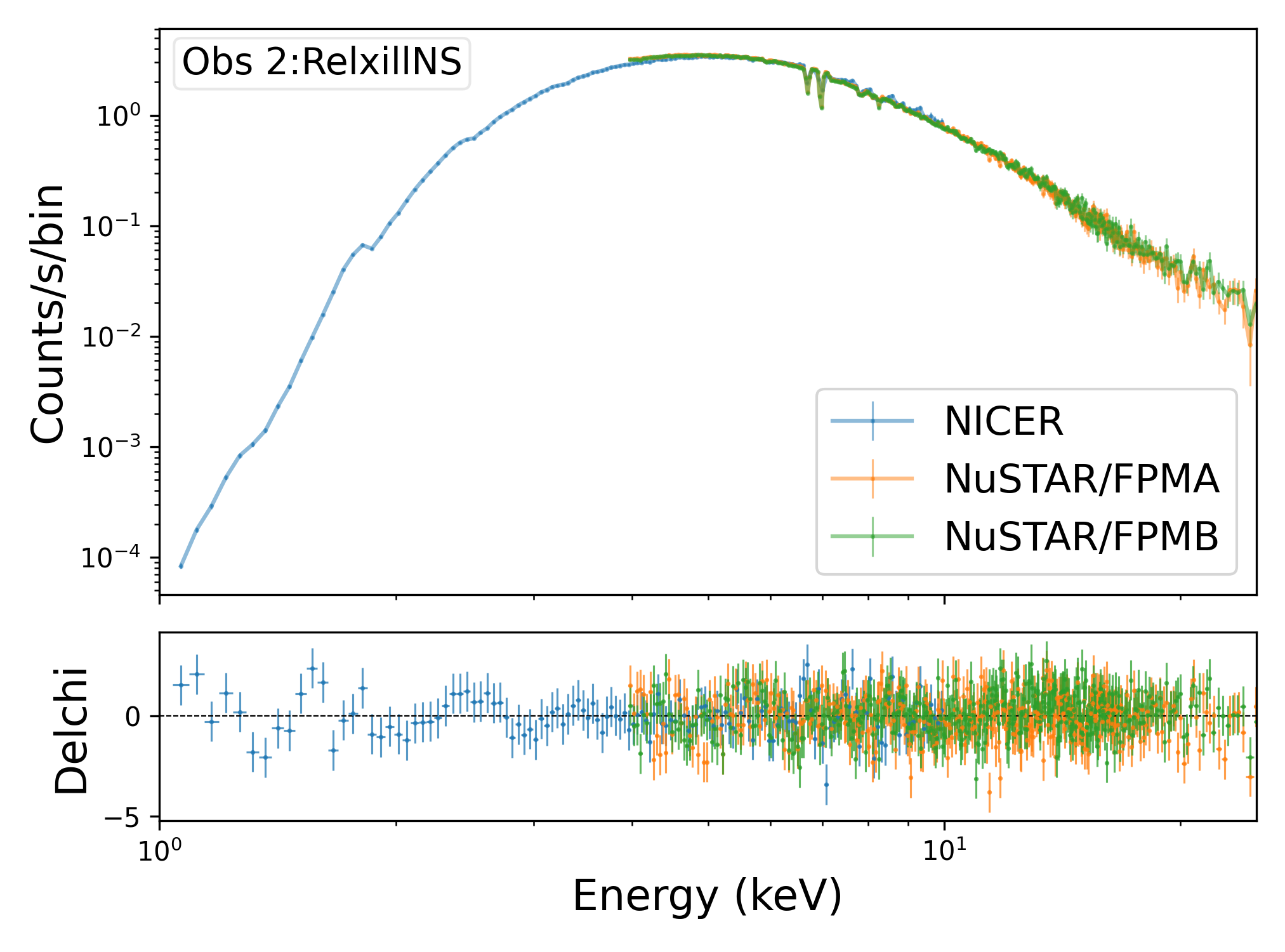}
	\vskip\baselineskip
	\centering
	\includegraphics[width=0.49\textwidth]{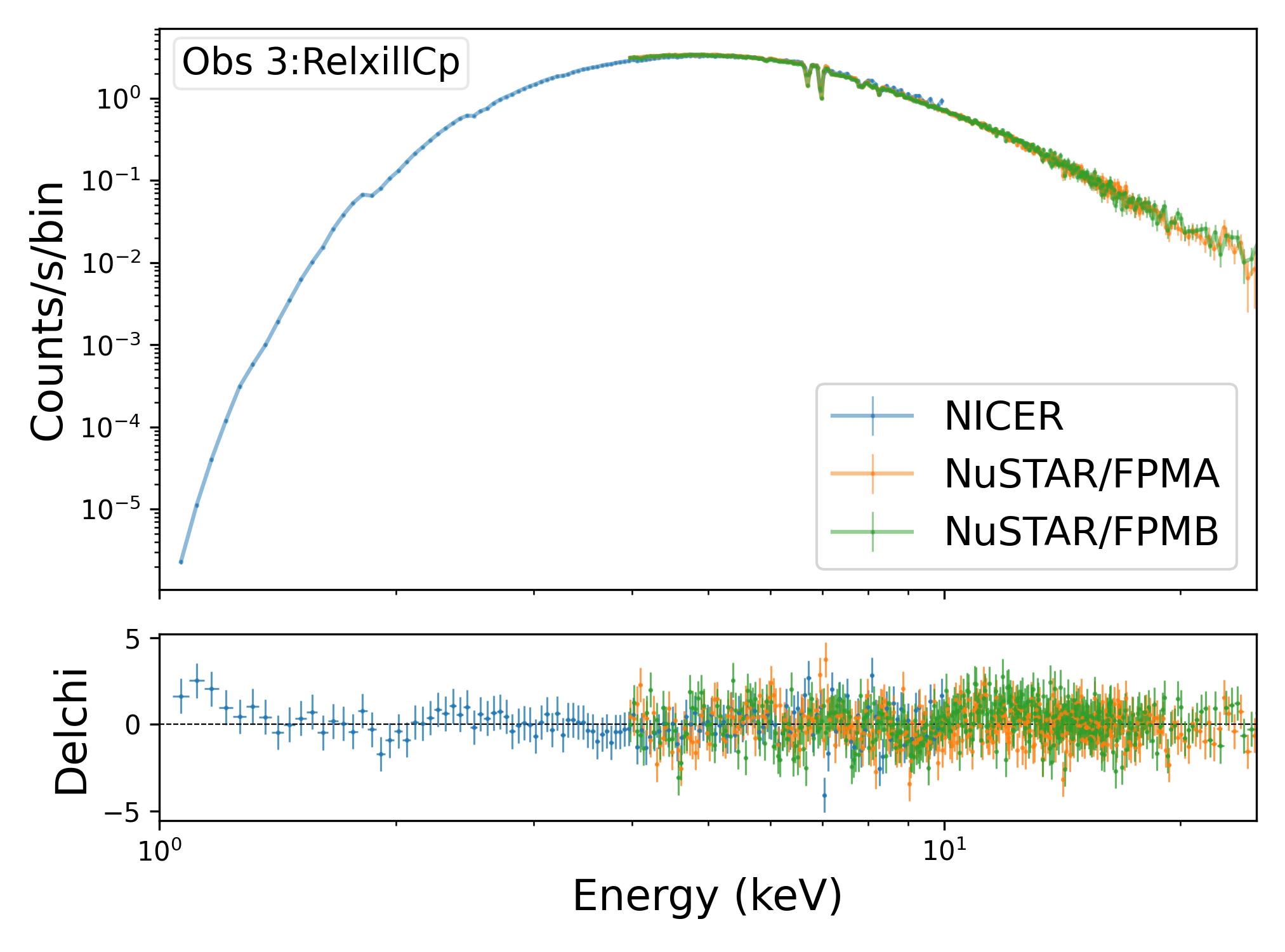}
	\hfill
	\centering
	\includegraphics[width=0.49\textwidth]{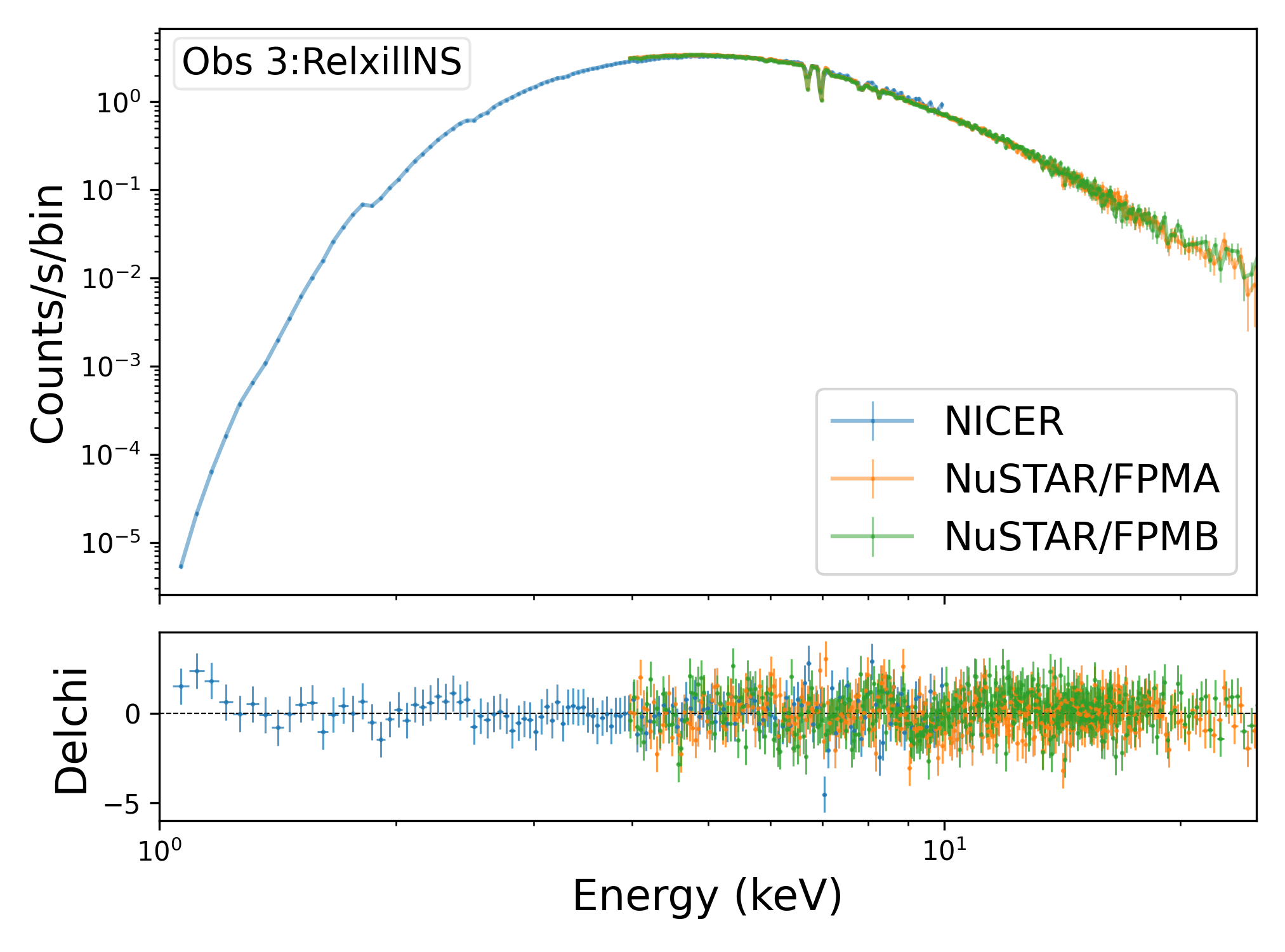}
	\caption{The model and residual plots (top and bottom panels of each subplot) for the three observations in three rows. The plots in the left column use \texttt{relxillCp}, and the plots in the right column use \texttt{relxillNS} to fit the data. We have used blue points to mark the \textit{NICER} data, orange for \textit{NuSTAR} FPMA, and green for \textit{NuSTAR} FPMB.
	}
	\label{fig:delchi}
\end{figure*}

The reflection component of \src{} in the soft state has been argued to be dominated by the returned disk emission \citep{Connors:2021nml}. The fact that we find only weak power-law emission appears to support this scenario. In this case, the incident spectrum illuminating the disk and producing the reflection component would be soft, making the use of the \texttt{relxillCp} model, which assumes a power-law input, inappropriate. To test how this could impact our spin measurements, we replace the \texttt{relxillCp} with \texttt{relxillNS}, which assumes the incident spectrum to be a single-temperature blackbody instead of a power-law. This model can mimic the effect of returning radiation in the soft state of black hole binaries by adopting a soft incident spectrum \citep[e.g.,][]{Connors2020ApJ...892...47C, Connors:2021nml}. It should be noted that the actual incident spectrum at each radius is the sum of returned blackbody spectra from all other radii, rather than a single-temperature blackbody. Modeling this effect self-consistently requires extensive ray tracing and radiative transfer calculations \citep[e.g.,][]{Mirzaev:2024qcu, Kourmpetis2026arXiv260114860K}. In this context, fitting the data with the \texttt{relxillNS} model can serve as a proof-of-principle test of the returning-radiation effect. If the source is in the hard state, where coronal emission dominates the spectrum, the returning-radiation effect can be accounted for using the \texttt{relxilllpCp} model. In this model, returning radiation is incorporated by modifying the emissivity profile for a lamppost coronal geometry without altering the incident spectrum \citep{Dauser2022MNRAS.514.3965D}. Simulations with self-consistent calculations have shown that this treatment is valid for highly ionized disks \citep{Mirzaev2024ApJ...965...66M}. However, such a model does not currently exist for the soft state in our case. Therefore, the \texttt{relxillNS} model provides a reasonable approximation among the available models and allows us to assess the systematics of different reflection models in spin measurements.

This alternative model also yielded an excellent fit to the data, with a $\chi^2$ comparable to or slightly better than that of the \texttt{relxillCp} model. The best-fit parameter values are listed in Table ~\ref{NS} and the residuals are shown in the right panels of Figure~\ref{fig:delchi}. With this model, the inclination measurements show a higher value (52--58 degrees). The spin measurement still yields high values, but the estimate does not indicate maximal spin. The choice of models seems to have minimal impact on the spin measurement. We have also used a lamppost variant of the reflection model \texttt{relxillLp}, to test our fit, but there is no significant improvement in the results. We have also repeated the analysis using both \texttt{relxillCp} and \texttt{relxillNS} models with modifications in $f_{col}$. There is a negligible effect on the spin estimates.

We again treat $M_{\rm BH}$ as an additional degree of freedom and observe an improvement in the $\chi^2$ values for the analysis with \texttt{relxillNS}. $M_{\rm BH}$ is measured to be 6.7--15 $M_{\odot}$. \texttt{relxillNS} is statistically preferred among the phenomenological models tested in this paper, but that is not the same as saying the physical origin of the disk illumination has been settled.

\subsection{Joint fitting}

By fitting the three observations separately, we find variabilities of parameters that should be constant (e.g., $M_{\rm BH}$ and $i$). Therefore, we run a joint fitting for the three observations, linking the parameters that are not supposed to change on short timescales, e.g., $M_{\rm BH}$, $i$, $a_*$, $A_{\rm Fe}$, column density of the \texttt{Tbabs} component, and calibration parameters. 

The best-fit parameters are shown in Table~\ref{CP_all} and ~\ref{NS_all} for \texttt{relxillCp}- and \texttt{relxillNS}-based models, respectively. The reduced-$\chi^2$ values are $3075.1/2627 \approx 1.17$ and $2768.4/2626 \approx 1.05$. For the \texttt{relxillCp} model configuration, we obtain $a_*=0.967_{-0.008}^{+0.004}$ and $M_{\rm BH}=9.14_{-0.39}^{+0.91}~\mathrm{M}_\odot$. For the \texttt{relxillNS} model configuration, we obtain $a_*=0.964_{-0.003}^{+0.006}$ and $M_{\rm BH}=12.68_{-0.86}^{+0.82}~\mathrm{M}_\odot$. When additional tests are performed by incorporating small fractional systematic uncertainties, the inferred errors increase noticeably. The resulting distributions become significantly broader. This behavior suggests that the $\chi^2$ joint-fit uncertainties primarily capture formal statistical errors. As a consequence, they are likely to underestimate the true level of uncertainty, particularly those arising from underlying systematic effects. Based on the mass estimation and distance value fixed at 10~kpc, we calculate the Eddington ratio to be $12.66_{-0.78}^{+0.86}$ \%, $12.09_{-0.73}^{+0.91}$ \% and $11.74_{-0.69}^{+0.89}$ \%, respectively, for the three observations. The inferred Eddington ratios ($\sim 12\%$) place the source well within the regime where the standard geometrically thin, optically thick accretion disk described by the Novikov–Thorne model is expected to be valid. This supports the application of thin-disk models in our analysis.

Once again, it appears that the choice of the reflection model affects the measurements of the black hole mass, while its impact on the spin is marginal. Notably, regardless of which reflection model is used, and whether separate or joint fitting is performed, a high black hole spin is consistently obtained for \src{}. 

\section{Spectral fitting with Bayesian statistics}
\label{s-bxa}

The application of Bayesian methods to the spectral analysis of XRBs has gained popularity in the scientific community due to its ability to incorporate prior knowledge of specific parameters and provide a more intuitive framework for probabilistic inference \citep{ge2022reconstructing,Witzel:2018kzq,Buchner:2014nha}. This section focuses on the basic formulation of the Bayesian approach, the selection of parameter priors, and the computational techniques and steps used in our study to derive posterior distributions. 

In contrast to the traditional frequentist approach, the Bayesian framework offers a more comprehensive view of parameter space. The model parameters are assumed to be unknown variables, and the probability distributions encode our state of knowledge \citep{box2011bayesian,buchner2023statistical}. At the core of Bayesian inference lies Bayes' theorem. It relates the posterior probabilities of the parameter set $\theta$ for model M to the product of the prior probability distributions of the parameters and the likelihood of the data. 

\begin{equation} \label{eq:Baye}
    P(\theta | D,M) = \frac{P(D | \theta,M) P(\theta|M)}{P(D|M)}
\end{equation}

Here, $P(D | \theta,M)$ is the likelihood of the data given the parameters of our model, $P(\theta | M)$ is the prior distribution encapsulating our prior knowledge or beliefs about $\theta$. $P(D|M)$ is the evidence, or marginal likelihood, defined as $P(D|M) = \int P(D | \theta,M) P(\theta|M) d\theta$. The posterior distribution $P(\theta | D,M)$ represents our updated knowledge about the parameters after analysis of the data D. For posterior sampling, we employ the nested sampling method using  BXA v4.1.1 and UltraNest v3.6.4, with PyXspec v2.1.4 and Python v3.9.19 \citep{buchner2016statistical,buchner2019collaborative,buchner2023nested,buchner2021ultranest,gordon2021pyxspec}. Nested sampling transforms a multi-dimensional evidence integral into a one-dimensional integral over the prior space. In each nested sampling iteration, it repeatedly replaces the worst-fitting point with a better one, gradually “nesting” into higher-likelihood regions.

In order to explore the parameter space and assess the robustness of our spectral fits, we perform a Bayesian analysis using the same models as discussed in the previous section. We call our model with \texttt{relxillCp} as \texttt{Model 0} and that with \texttt{relxillNS} as \texttt{Model 1}. The models have many parameters, and we assign priors to each of them in suitable ranges. Our analysis considers Gaussian priors for $\dot{M}_{\rm BH}$ in \texttt{kerrbb} and for the parameters in \texttt{TBabs} and \texttt{pcfabs} and Jeffreys priors for $norm$ in \texttt{relxill}. For the rest of the parameters, we assign them uniform priors. The remaining model parameters, such as the parameters in \texttt{mbpo}, \texttt{gauss}, and \texttt{xstar}, are fixed to the best-fit values derived from the $\chi^2$ analysis. Prior types and ranges for all the free parameters are tabulated in Table~\ref {priorCP} and Table~\ref {priorNS} for \texttt{Model 0} and \texttt{Model 1}, respectively. The likelihood function for the observed X-ray spectrum is modeled as a Poisson distribution. 

\begin{figure} [h]
    \centering
    \includegraphics[width=\linewidth]{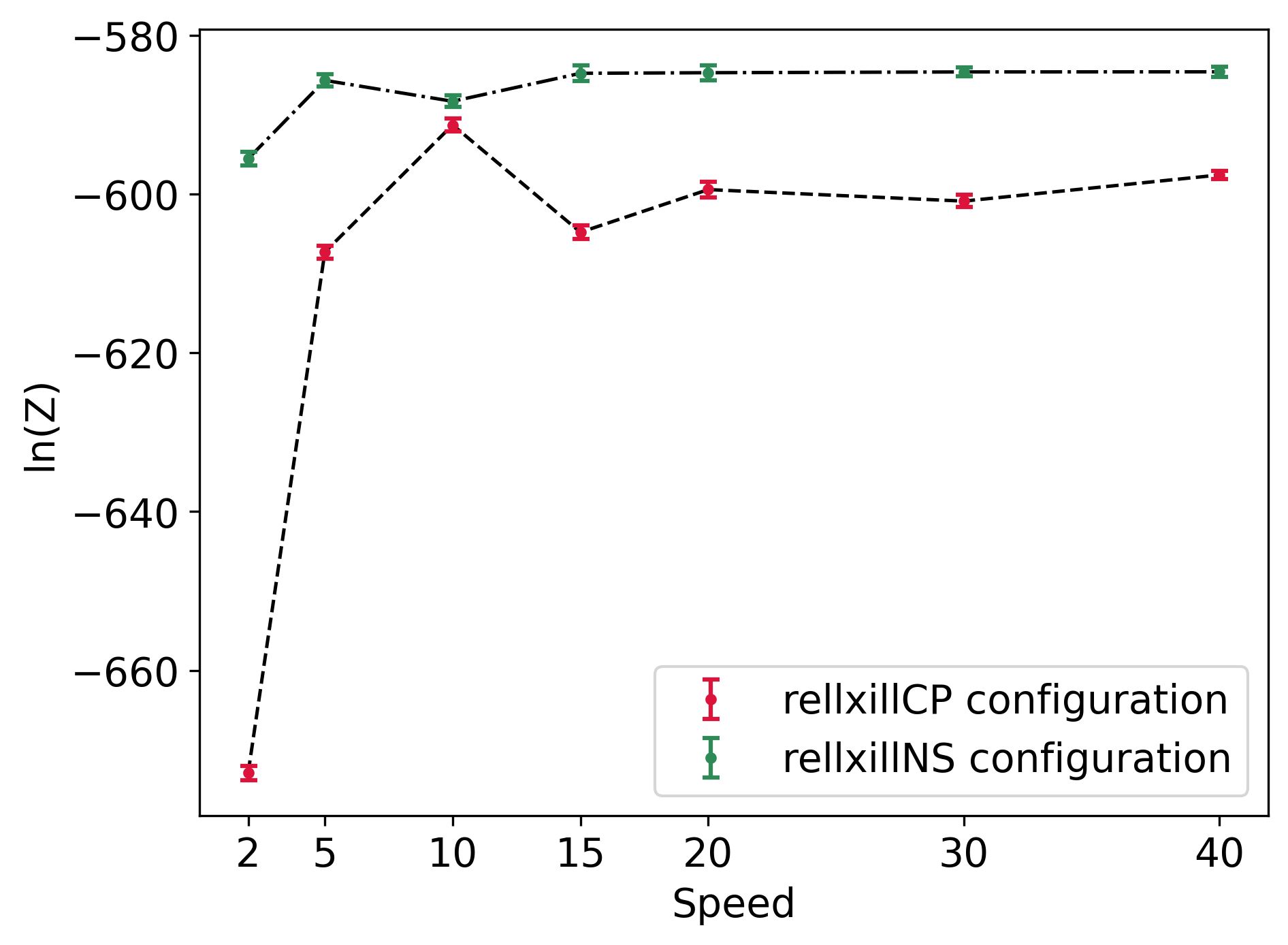}
    \caption{Evolution and convergence of $\ln (Z)$ for different values of nested sampling \texttt{speed} parameter for the analysis of Obs~1.}
    \label{fig:logZ}
\end{figure}

To ensure reliable convergence in Bayesian parameter estimation, we first run test cases of the BXA pipeline on Obs 1 with different \texttt{speed} settings. This parameter in BXA controls the trade-off between sampling efficiency and computational cost in nested sampling \citep{buchner2024relative}. A lower \texttt{speed} value leads to faster convergence by reducing the number of nested sampling steps per likelihood evaluation, but may result in less accurate evidence estimates. Conversely, a higher \texttt{speed} improves sampling thoroughness at the cost of longer runtime. We experiment with both lower and higher \texttt{speed} values, examining the resulting posterior distributions, evidence values, and live point evolution to evaluate sampling performance. Convergence is assessed by visual inspection of trace plots, the stabilization of log-evidence values, and the consistency of parameter posteriors across runs. Fig.~\ref{fig:logZ} shows the natural logarithm of the Bayesian evidence ($\ln (Z)$) values for each \texttt{speed} at which we ran the BXA analysis. We see a rise in the $\ln (Z)$ value as the \texttt{speed} increases, followed by a stabilization. Based on this, we select an optimal speed setting that balances computational cost with sufficient sampling accuracy for the final analysis.

\begin{figure} [h]
    \centering
    \includegraphics[width=\linewidth]{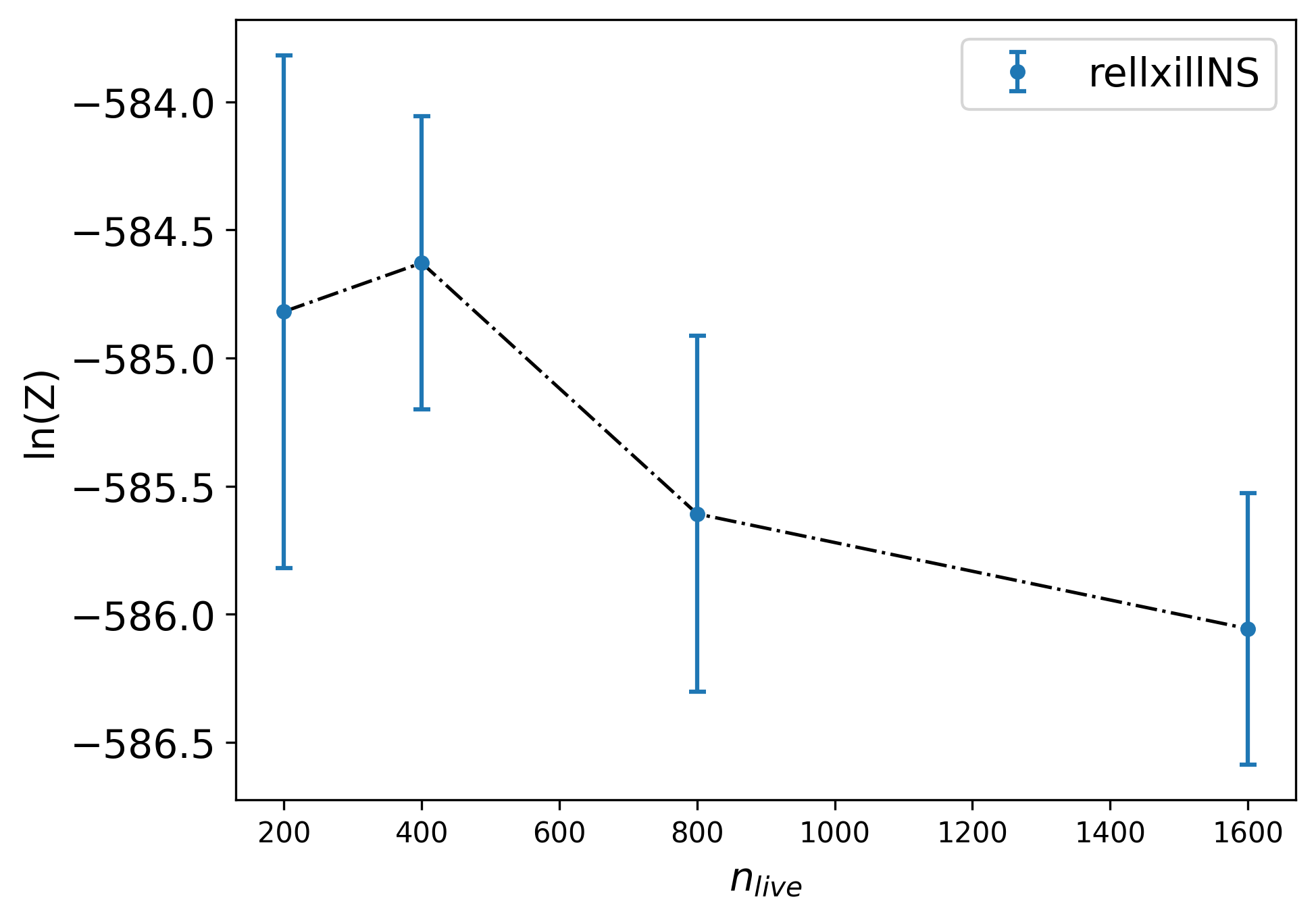}
    \caption{Evolution of $\ln (Z)$ for different values of the number of live points used in nested sampling for the analysis of Obs~1 using \texttt{Model 1}.}
    \label{fig:nlive}
\end{figure}

We conduct another test to understand the intricacies of the Bayesian parameter estimation by analyzing Obs 1 with \texttt{Model 1} setting  \texttt{speed}=30 and varying the number of live points $n_{live}$. We perform this test with \texttt{Model 1}, i.e., the model with \texttt{relxillNS} configuration, since it yields better $\ln (Z)$ values in the previous test. The \texttt{speed} is set to 30 since we consider it to be a decent choice for sampling efficiency based on the convergence of $\ln (Z)$ in Fig.~\ref{fig:logZ} and considering the computation time as we vary $n_{live}$. In nested sampling, $n_{live}$ is a critical parameter that dictates the trade-off between accuracy and computational cost \citep{chen2019improving, dittmann2024notes}. Increasing the number of live points provides a finer resolution of the prior volume, reducing the statistical error (which scales approximately as $1/ \sqrt{n_{live}}$) in the estimation of the prior volume "shells" at each iteration. It helps to robustly explore complex or multimodal posterior distributions, reducing the chance of "mode die-off," where an important peak in the likelihood is missed entirely. The number of likelihood evaluations, and thus the runtime, scales linearly with the number of live points. A higher $n_{live}$ means that each step involves more work, and more steps are needed to reach convergence, thereby increasing the overall computation time \citep{chen2019improving}. Fig.~\ref{fig:nlive} shows the $\ln (Z)$ for different $n_{live}$ values. In general, one can argue that the decrease in the $\ln (Z)$ value with an increase in $n_{live}$ is due to an overestimate produced by a run with too few live points that inadequately explored the parameter space. The difference in $\ln (Z)$ values plotted in Fig.~\ref{fig:nlive} is simply a statistical fluctuation within the error bounds.

\begin{figure*}[ht]
    \centering
    \includegraphics[width=0.49\textwidth]{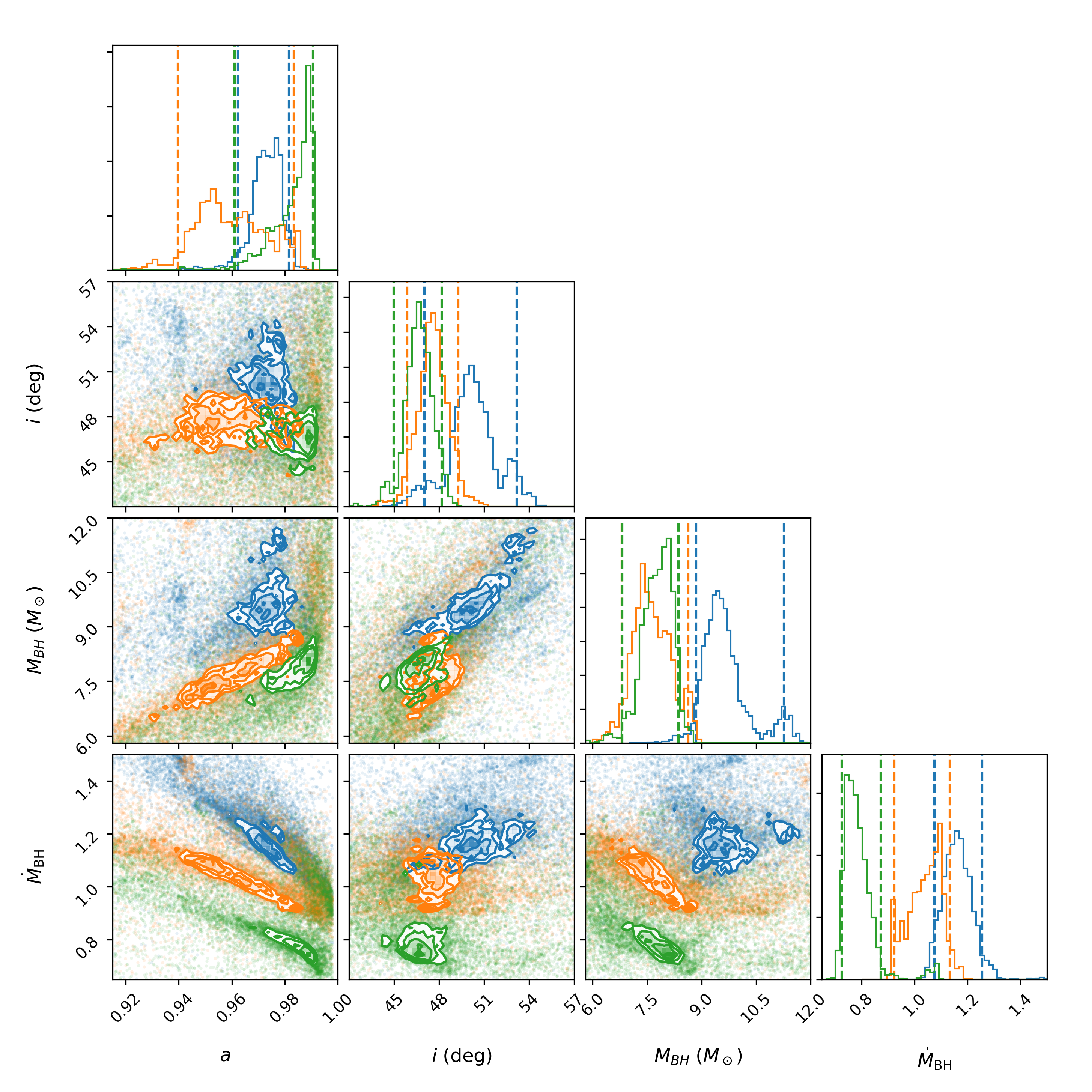}
    \centering
    \includegraphics[width=0.49\textwidth]{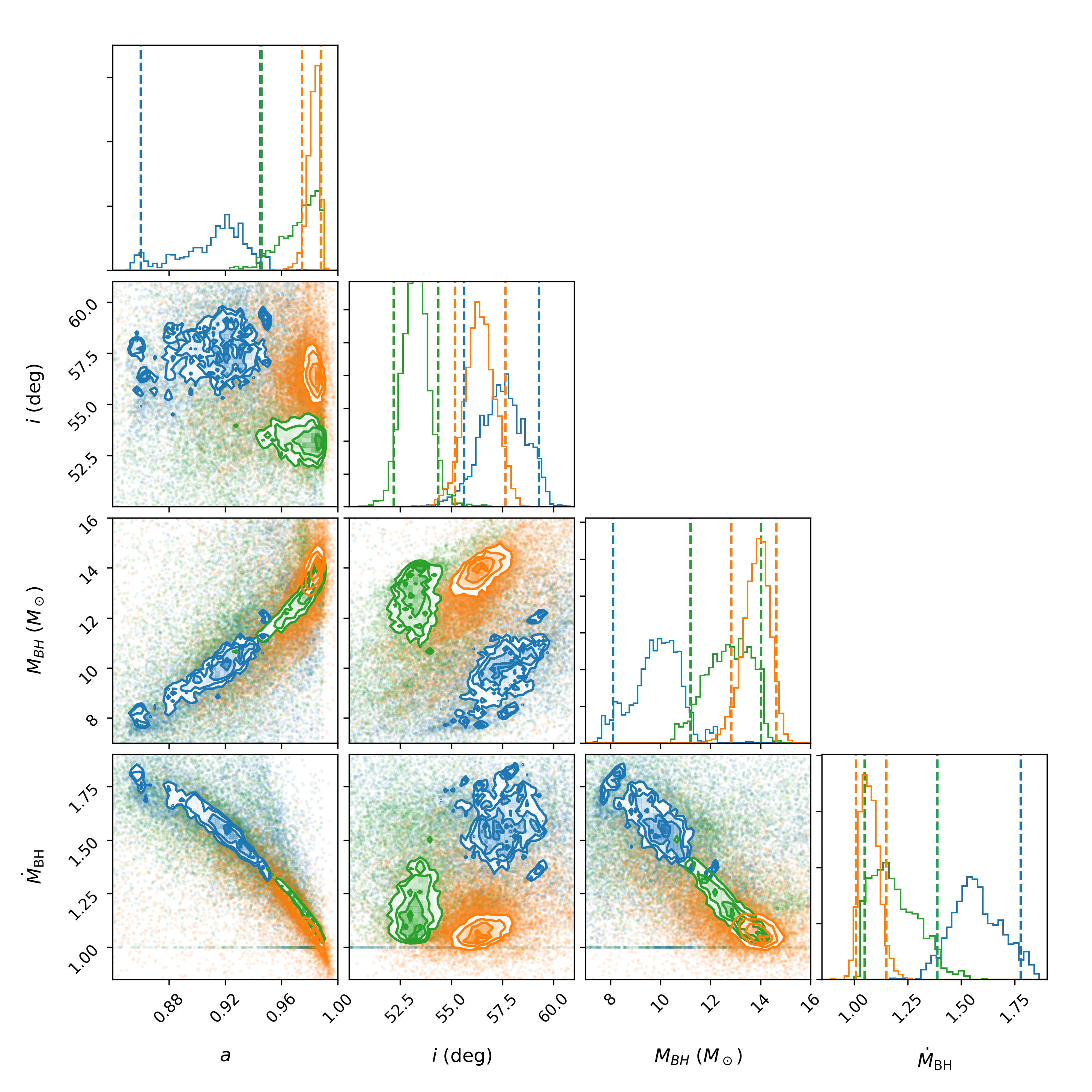}    
    \caption{Corner plots showing the parameters $a$, $i$, $M_{\rm BH}$, and $\dot{M}_{\rm BH}$. The left panel shows the posteriors for Obs~1 (blue), Obs~2 (orange), and Obs~3 (green) for the \texttt{relxillCp} model configuration. The right panel uses the same color scheme and displays the results from \texttt{relxillNS} model configuration. The full corner plots of each of the analyses are shown in  Fig.~\ref{fig:1C40}-~\ref{fig:3N40}. }
    \label{fig:corners}
\end{figure*}

When we run the full BXA analysis on all three observations, using \texttt{Model 0} and \texttt{Model 1}, each observation data set is treated independently with a similar choice of prior distributions as used while analyzing Obs~1. The nested sampling \texttt{speed} is set to 40, and $n_{live}$ is set to 400. The estimated posterior values from the Bayesian analysis of the 3 observations with $90\%$ intervals are summarized in Table~\ref{bxaCP} and ~\ref{bxaNS} for the 2 model configurations, respectively. Figs.~\ref{fig:1C40}-~\ref{fig:3N40} show the corner plots of the three observations with all the free parameters. The Bayesian evidence ($\ln (Z)$) is particularly useful for comparing different models. Since we have two models, we can calculate the evidence for each and compare them. The model with the stronger evidence is considered more likely to have generated the observed data. The curves in Fig.~\ref{fig:logZ} show that \texttt{Model 1} explains the data better than \texttt{Model 0}. The $\ln (Z)$ values for the analysis of the other observations also suggest \texttt{Model 1} being the preferred model. The evidence values are jotted down in Tables~\ref{bxaCP} and ~\ref{bxaNS}. The $L/L_{Edd}$ ratio for the 3 observations are $16.18_{-1.94}^{+3.51}$ \%, $11.05_{-1.98}^{+2.46}$ \% and $11.66_{-1.23}^{+1.36}$ \%  when evaluated using the posterior samples from the Bayesian analysis using \texttt{Model 1}. The inferred Eddington ratios ($\sim 11-16\%$) once again place the source within the standard geometrically thin, optically thick accretion disk regime described by the Novikov–Thorne.

By multiplying the individual posterior probability densities, we construct the final joint posterior, which provides a more robust estimate of the system properties. From this combined posterior, we infer a black hole spin $a_*= 0.96_{-0.03}^{+0.02}$, mass $M_{\rm BH} = 12.19_{-1.46}^{+1.22} \, M_\odot$, and inclination angle $i=55.75_{-1.37}^{+1.60}$ for \texttt{Model 1}. Combining the posteriors obtained when \texttt{Model 0} was used provides $a_*= 0.97_{-0.02}^{+0.02}$. In this case, $i=48.17_{-1.75}^{+1.93}$, and $M_{\rm BH} = 8.31_{-0.77}^{+0.93} \, M_\odot$. The improved constraints reflect the complementary information encoded in each observation and strengthen the reliability of our physical interpretation of the system. 

Based on the estimates of the BH spin, BH mass, and mass accretion rate obtained with \texttt{Model 1}, the inner disk temperature lies in the range $\sim 1$--$1.6$~keV (see Tab.~\ref{bxaNS}). We also note that the best-fit values of $kT_{\rm bb}$ in \texttt{relxillNS} do not track the inner disk temperatures inferred from the thermal continuum component. As explained in Sec.~\ref{subsec:impact}, \texttt{relxillNS} mimics the effect of returning radiation in the soft state of BH binaries by assuming the incident spectrum over the entire disk to be a single-temperature blackbody. The parameter $kT_{\rm bb}$ should not be interpreted as a direct measure of the observed inner disk temperature, because the incident spectrum at each radius deviates from a blackbody due to Doppler shifting/beaming, gravitational redshift, and the superposition of radiation from other radii with different temperatures.

We focus on the correlations among $a_*$, $i$, $M_{\rm BH}$, and $\dot{M}_{\rm BH}$. Spin estimates exhibit a positive correlation with $M_{\rm BH}$ and a negative correlation with $\dot{M}_{\rm BH}$ in almost all analyses. Positive correlations between $M_{\rm BH}$ and $i$ are also evident. We also observe a negative correlation between $M_{\rm BH}$ and $\dot{M}_{\rm BH}$. These can be easily visualized from the corner plots that contain only these parameters in Fig.~\ref{fig:corners}. Apart from the degeneracies among these parameters, posterior distributions from some analyses indicate the presence of secondary modes. This feature is particularly noticeable in specific observations. While \texttt{Model 1} infers posterior distributions of $a_*$, $i$, $M_{\rm BH}$, and $\dot{M}_{\rm BH}$, which are largely well-behaved and approximately Gaussian distributions, some of the posteriors obtained via \texttt{Model 0} deviate from Gaussian-like distributions. Notable examples include the $M_{\rm BH}$ and $i$ posteriors of Obs 2 and $a_*$ and $\dot{M}_{\rm BH}$ for Obs 3. The only multimodal feature in the distribution obtained via \texttt{Model 1} is observed in the $a_*$ posterior.

Talking of the reflection and \texttt{simplcutx} parameters, Bayesian runs with both \texttt{Model 0} and \texttt{Model 1} fetch similar estimates for some parameters like $f_{\rm sc}$, $\Gamma$, $kT_{bb}$, $A_{\rm Fe}$, and $log(\xi)$. The $A_{\rm Fe}$ estimates are slightly higher with \texttt{Model 0}. The estimates of the \texttt{TBabs} and \texttt{pcfabs} parameters show a significant match with the analysis results from the \texttt{XSPEC} fit. A negative correlation between the $N_{\rm H}$ of \texttt{TBabs} and $CvrFract$ of \texttt{pcfabs} is noticed in all the corner plots. 

Going through Fig.~\ref{fig:1C40}-~\ref{fig:3N40}, we observe some more degeneracies based on the correlation plots between pairs of parameters. In addition to the negative correlation between the $N_{\rm H}$ of \texttt{TBabs} and $CvrFract$ of \texttt{pcfabs} which is present in all the corner plots, $N_{\rm H}$ of \texttt{TBabs} and $N_{\rm H}$ of \texttt{pcfabs} also show negative correlation when the data is analyzed with \texttt{Model1}. $f_{\rm sc}$ and $\Gamma$ also exhibit strong positive correlation in the corner plots in Fig.~\ref{fig:1C40}-~\ref{fig:3N40}. The $norm$ posteriors of the respective reflection components in both model configurations display some signatures of negative correlation with some other parameters of the respective reflection models, like the $log(n_{e})$ in Fig.~\ref{fig:1C40}, ~\ref{fig:2C40} and ~\ref{fig:3C40} which are obtained via \texttt{Model 0} and $log(\xi)$ and $kT_{bb}$ in Fig.~\ref{fig:1N40}, ~\ref{fig:2N40} and ~\ref{fig:3N40} obtained via \texttt{Model 1}.

\section{Discussion}\label{s-dis}

By analyzing three observations of 4U~1630--47 during its 2022 outburst using \textit{NICER} and \textit{NuSTAR} data, we are able to constrain its BH spin and other important parameters. Spin measurements from $\chi^2$ statistics joint fitting shows $a_*= 0.967_{-0.008}^{+0.004}$ and $a_*= 0.964_{-0.003}^{+0.006}$ for the \texttt{relxillCp} and \texttt{relxillNS} model configurations.

Spin measurements from Bayesian analysis reports $a_*= 0.97_{-0.02}^{+0.02}$ and $a_*= 0.96_{-0.03}^{+0.02}$ for \texttt{relxillCp} and \texttt{relxillNS} model configurations. These values correspond to the highest-evidence configuration in the Bayesian analysis framework. The nested-sampling settings, namely \texttt{speed} and $n_{live}$ values, are sufficient to ensure convergence based on the analysis done to study the $\ln (Z)$ convergence in Section~\ref{s-bxa}. Similar $\ln (Z)$ convergence tests have been carried out in a previous study~\citep{buchner2024relative}. The Bayesian intervals provide a more realistic uncertainty estimate in comparison to the $\chi^2$ joint-fit, which captures formal statistical errors and underestimates underlying systematic effects.

Bayesian analysis helps us to distinguish which model is better suited to describe the data when we compare the Bayesian evidence values for the two model configurations. Higher evidence of the Bayesian fit with \texttt{relxillNS} model hints at the potential presence of returning radiation in our system. \texttt{relxillNS} does not constitute a fully self-consistent model for returning radiation. Posterior distributions of certain parameters tend to accumulate near the edges of their allowed prior ranges. This behavior reflects underlying limitations in the model. In addition, the role of high-density reflection has not been comprehensively investigated within the present framework. The photon index $\Gamma$ frequently pegs at its upper limit of 3.5, and the disk electron density $log(n_e)$ hits the lower boundary of 15 in several relxillNS fits. This is noticed for both $\chi^2$ and Bayesian statistics. Widening the upper bound on $\Gamma$ has a negligible impact. Similarly, $log(n_e)$ pegging at 15 may signal that the data prefer a lower density, with physical implications for the reflection spectrum. From our analysis, it is evident that a softer illumination profile yields a statistically improved fit to the data. However, the physical origin of illumination remains unclear. A more detailed exploration of this aspect is therefore warranted in future work.

We acknowledge that the modeling of the photoionized absorber in \src{} as a single-zone XSTAR component is a simplification. Previous studies, particularly those using high-resolution spectroscopy (e.g., \citealt{Rozanska:2013foa,King:2014sja,Parra:2025oaj,Parra:2023obj,Fan:2024mhe}), indicate that \src{} often features complex, multi-phase disk winds. While our custom photoionization table successfully models the prominent Fe XXV and Fe XXVI absorption features in the NICER and NuSTAR spectra, a single-zone approach may not capture the full, multi-layered nature of the wind. Furthermore, we did not apply explicit independent velocity-shift constraints to the individual Fe lines. These simplifications, particularly the assumption of a single zone, may introduce systematic degeneracies. Because the wind absorption features exist mainly in the 6–8 keV range, partially overlapping with the iron line and the edge of the relativistic reflection component, a simplified wind model might lead to degeneracies, particularly affecting the ionization parameter and iron abundance of the reflection model. However, testing different turbulent velocities showed that our main conclusions regarding the reflection parameters and the presence of both features hold. We have also noted that the impact on the BH spin estimate is negligible. Hence, the current approach provides a robust description of the spectral behavior. Moreover, a fixed emissivity $q=3$ may underestimate the steepness of the profile for a rapidly spinning BH. Hence, we carry out some tests with $q=5$ and $q=7$ on Obs 1. Our tests show that the value of $a_*$ varies from $0.983_{-0.009}^{p}$ for $q=3$ to $0.948_{-0.012}^{+0.015}$ for $q=5$ and to $0.911_{-0.005}^{+0.01}$ for $q=7$, when analyzed with the \texttt{relxillCp} model. For the \texttt{relxillNS} model, the inferred $a_*$ are $0.891_{-0.014}^{+0.013}$ and $0.878_{-0.005}^{+0.016}$ for $q=5$ and $q=7$, respectively, which was $0.928_{-0.022}^{+0.022}$ for $q=3$. The best-fit values are displayed in Tables~\ref{qCP} and ~\ref{qNS}. $\Delta a^* \approx 0.035 - 0.072$ and $\Delta a^* \approx 0.037 - 0.05$, which exceeds the quoted statistical errors ($ \pm 0.01–0.03$). Therefore, the emissivity assumption actually dominates the error budget compared to the statistical intervals. That being said, the high spin nature of \src{} remains a robust conclusion. 

For \src{}, \citet{Liu:2021tyw} find the following estimates of the black hole spin parameter  $a_* = 0.817_{-0.014}^{+0.014}$ from the 2020 outburst  Insight-HXMT observations. These results are produced by the analysis of the reflection features.  The source was in the intermediate state. The model used to analyze the source was \texttt{TBabs}*(\texttt{diskbb}+\texttt{relxillCp}). From the \textit{NuSTAR} observation on 9 March 2013 \citet{King:2013fma} find $a_* = 0.985_{-0.014}^{+0.005}$ employing the reflection model \texttt{refbhb} \citep{Ross:2007qk,Reis:2008ja}, which is designed for modeling the reflection spectrum of Galactic BHs in the soft or intermediate states, when the thermal soft X-ray photons of the accretion disk can contribute to the reflection spectrum. With the same \textit{NuSTAR} data and employing the lamppost version of {\tt relxillCp}, \citet{Tripathi:2020yts} still find a very high black hole spin parameter, $a_* = 0.99_{-0.005}^{+0.003}$. From the analysis of AstroSat and Chandra data, \citep{Pahari:2018toe} reports the BH spin parameter $a_* = 0.92_{-0.04}^{+0.04}$ from the analysis of the thermal spectrum of the disk. \citet{Kushwaha2023MNRAS.524L..15K} finds the spin $a_* = 0.92_{-0.02}^{+0.02}$ using \textit{NICER} and IXPE data for the 2022 outburst. These spin estimates are obtained using the continuum-fitting method. Compared to earlier analyses of this source, most of which relied on either reflection or continuum methods in isolation, we emphasize the improved constraints achieved through combining the two methods. We have confirmed that the black hole in this source has a high spin. In another study on understanding the distribution of BH spins in X-ray binaries \citep{Draghis:2025izq}, the spin estimate of our source is $0.86_{-0.21}^{+0.10}$.

By combining the reflection and continuum-fitting methods, we can also constrain the BH mass and inclination. The preferred model (\texttt{relxillNS}) provides a $M_{\rm BH} = 12.68_{-0.86}^{+0.82} \, M_\odot$ and $i=56.92_{-0.52}^{+0.74}$ degrees with $\chi^2$ fitting. When analyzed in the Bayesian framework, the mass and inclination estimates are $M_{\rm BH} = 12.19_{-1.46}^{+1.22} \, M_\odot$ and $i=55.75_{-1.37}^{+1.60}$ degrees. 

A key advantage of our Bayesian treatment is the ability to fully capture parameter degeneracies and propagate uncertainties systematically. The posterior distributions from the Bayesian analysis reveal the extent of the degeneracies between them. Older studies that made the estimations of inclination, and mass of the BH in the binary system report  $i= 60^\circ-70^\circ$ and $M_{\rm BH} = 10.0_{-0.1}^{+0.1} \mathrm{M}_\odot$ \citep{Kuulkers:1997fh,Seifina:2014ura}. The detection of a dip in the X-ray light curve during the 1996 outburst \citep{Kuulkers:1997fh} indicates that the inclination of the system is fairly high in the range $60^\circ-75^\circ$. The scaling of the correlation between the photon index of the Comptonized spectral component and low-frequency QPOs and the mass accretion rate was applied to the BH mass and inclination angle estimates. The scaling technique, which relies on XTE J1550-564, GRO 1655-40, and H1743-322 as reference sources, allows the evaluation of a black hole mass to be $10.0_{-0.1}^{+0.1} \mathrm{M}_\odot$ and puts a constraint on the inclination ($i \leq 70^\circ$) in 4U 1630--47 \citep{Seifina:2014ura}. Other studies use these values in the course of their projects on 4U~1630--47 \citep{Pahari:2018toe,Kushwaha2023MNRAS.524L..15K}. About the inclination discrepancy, the tension between our results ($i \simeq 50^\circ-60^\circ$) and the dip-inferred values ($i > 60^\circ$) suggests that current reflection models may be affected by the complex disk wind known in this source. While testing our parameter estimation with emissivity index $q=5$ and $q=7$, we have noticed that the higher the inferred inclination, the lower the inferred spin estimate. The mass estimate may also be affected by the same inclination-related systematics.

Lastly, our study demonstrates the strength of Bayesian analysis tools in the context of high-dimensional spectral modeling. In recent years, several studies have used the strong Bayesian framework to study X-ray sources \citep{ge2022reconstructing,Witzel:2018kzq, Dias:2024ezj}. The X-ray binary MAXI J1820+070 during 'hard' states of its 2018-2019 outburst has been studied extensively using Bayesian analysis of observations from XMM-Newton and \textit{NuSTAR}. They have estimated the spin of MAXI J1820+070 in the Bayesian framework \citep{Dias:2024ezj}. Bayesian methods are used for reconstructing the parent distributions of X-ray spectral parameters of active galactic nuclei in large surveys \citep{ge2022reconstructing}. The variability of supermassive BHs is being studied by Bayesian computational analysis \citep{Witzel:2018kzq}. Other work includes X-ray spectral modeling of active galactic nuclei using a Bayesian framework for model comparison and parameter estimation of X-ray spectra \citep{Buchner:2014nha}. The Bayesian perspective prioritizes uncertainty quantification and statistical rigor essential for scientific inference in astronomy. Nevertheless, it is important to recognize that systematic effects can lead to underestimated uncertainties in both Bayesian and frequentist frameworks. Sources of systematics include assumptions in spectral models, calibration uncertainties, and the choice of prior distributions. These limitations remind us that seemingly tight constraints may mask underlying model dependencies. Future work could address these issues. Such efforts will further strengthen the reliability of BH spin and parameter measurements. As X-ray instrumentation improves and data quality increases, we anticipate that such methods will play a central role in precision astrophysics, particularly in efforts to test strong-field gravity and probe accretion physics.

{\bf Acknowledgments --} This work was supported by the National Natural Science Foundation of China (NSFC), Grant No.~W2531002. D.D. also acknowledges support from the China Scholarship Council (CSC), Grant No. 2022GXZ005434. D.D. expresses his gratitude to Gitika Mall for insightful discussions that shaped the progress of this work in the initial phase. We are also sincerely thankful to Matteo Guainazzi, whom we had the pleasure of meeting during the I-HOW COSPAR Workshop 2024: A New Era of High-Resolution X-Ray Spectroscopy hosted by our group. His willingness to engage and expert guidance provided clarity and direction, which greatly benefited this project.

\bibliographystyle{apj}
\bibliography{ref}

\appendix
\section{The best-fit tables from Frequentist and Bayesian parameter estimations \& Corner Plots}

Here, we present the detailed results of both the frequentist and Bayesian analyses performed in this work, together with the posterior distributions obtained via Bayesian inference.

Tables~\ref{CP}-~\ref{NS_all} list the best-fit parameter values obtained from the XSPEC spectral fitting using the two \texttt{relxill} models, \texttt{relxillCp} and \texttt{relxillNS}. Tables \ref{CP} and \ref{NS} present the results from the fitting of the individual observations, while Tables \ref{CP_all} and \ref{NS_all} show the results obtained from the joint fitting of the observations with \texttt{relxillCp} and \texttt{relxillNS} respectively.

\begin{figure*} [hb]
    \centering
    \includegraphics[width=\linewidth]{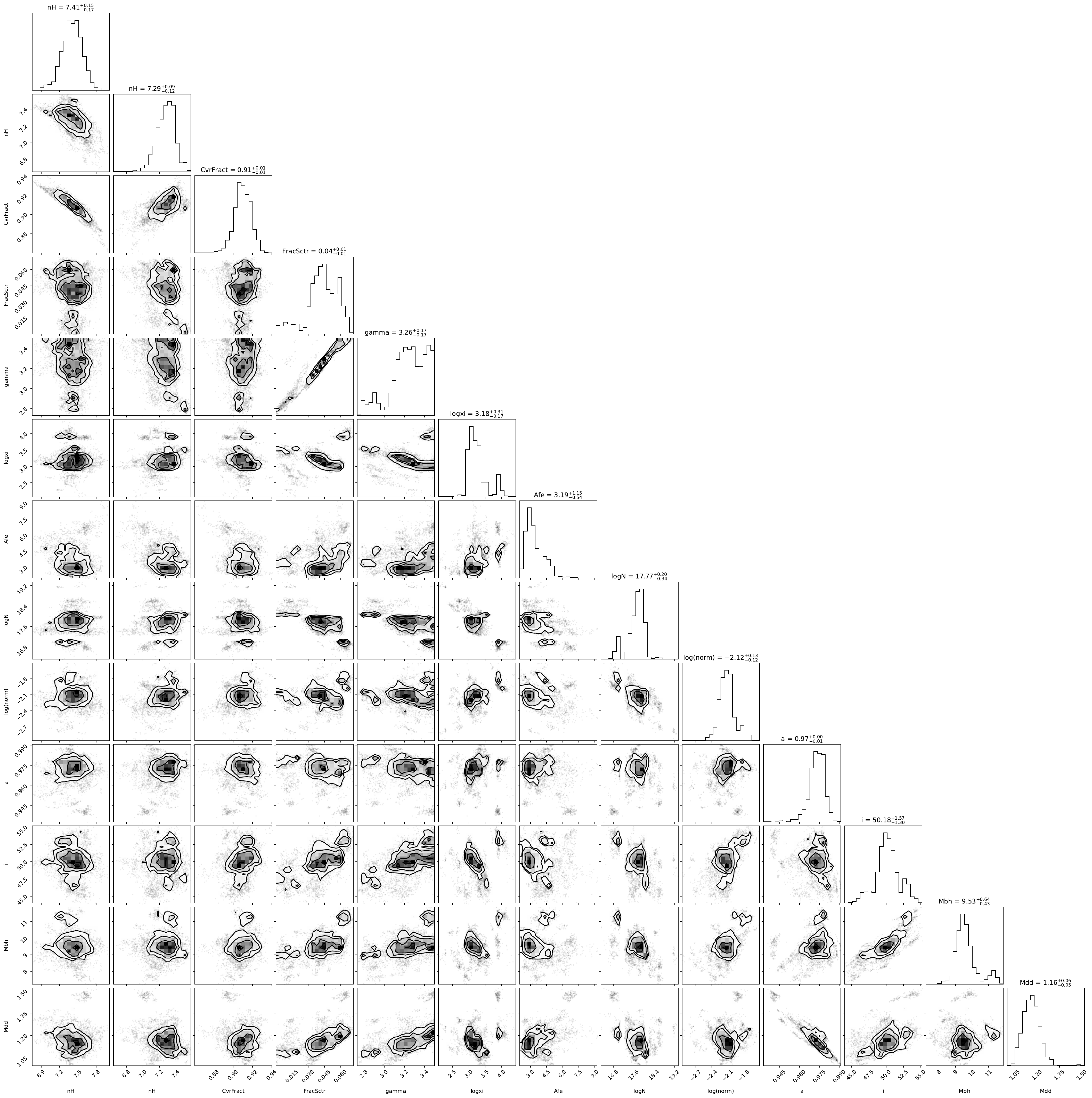}
    \caption{Corner plot of the Bayesian parameter estimation of Obs 1 using \texttt{Model 0}. The error values of each of the parameters correspond to $1\sigma$ of the respective posterior distributions.}
    \label{fig:1C40}
\end{figure*}

\begin{figure*} [h]
    \centering
    \includegraphics[width=\linewidth]{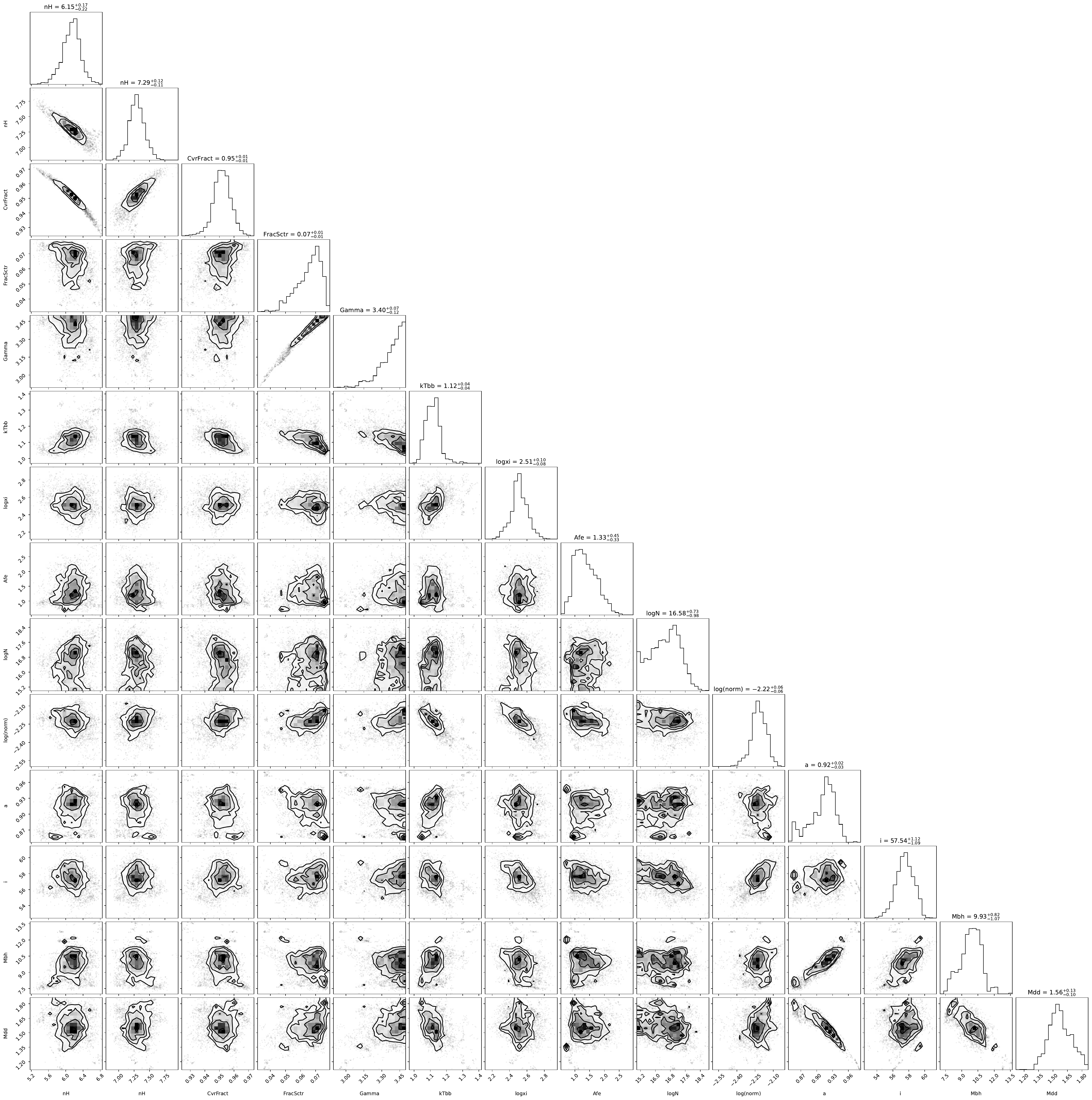}
    \caption{Corner plot of the Bayesian parameter estimation of Obs 1 using \texttt{Model 1}. The error values of each of the parameters correspond to $1\sigma$ of the respective posterior distributions.}
    \label{fig:1N40}
\end{figure*}

\begin{figure*} [h]
    \centering
    \includegraphics[width=\linewidth]{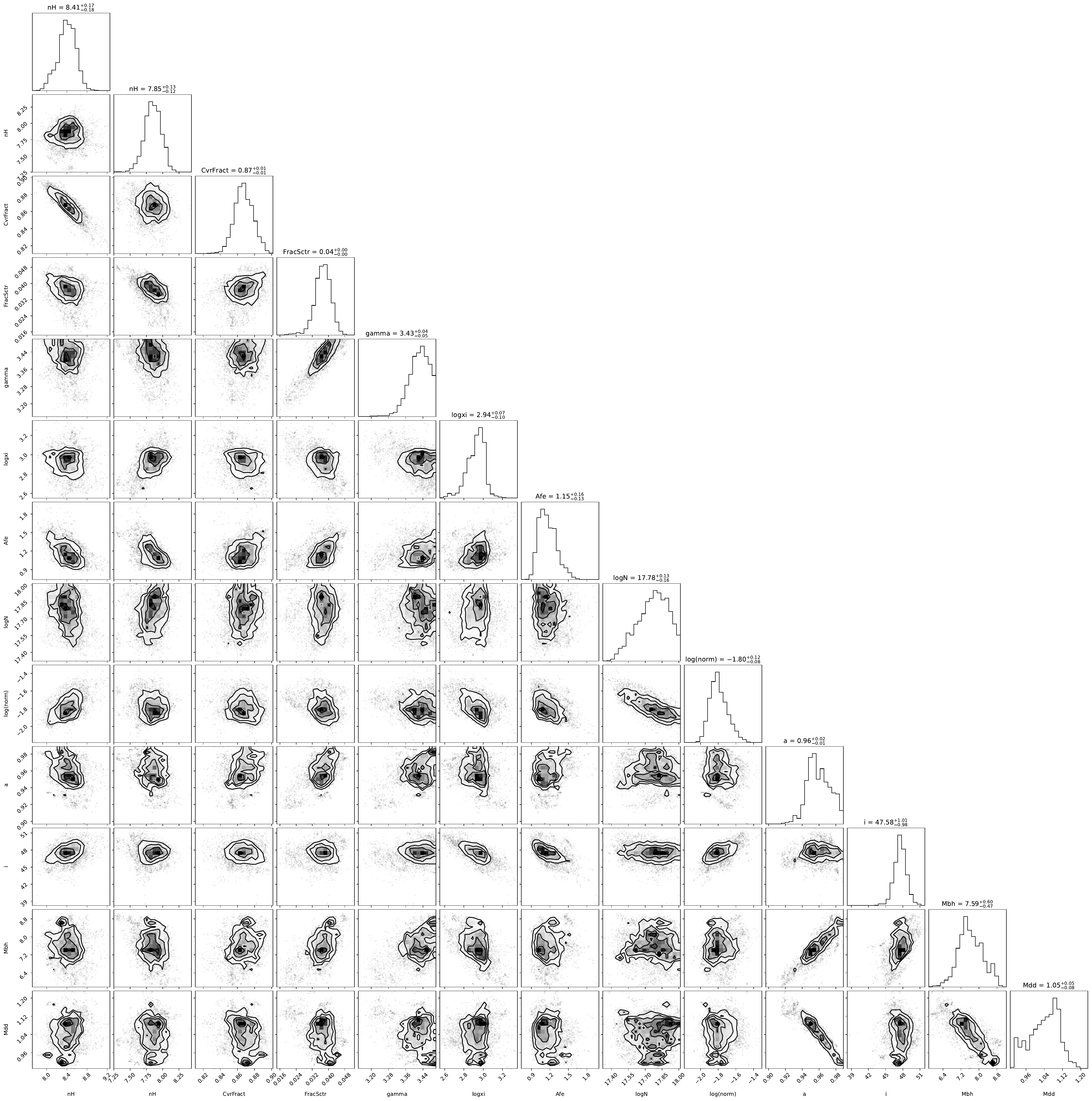}
    \caption{Corner plot of the Bayesian parameter estimation of Obs 2 using \texttt{Model 0}. The error values of each of the parameters correspond to $1\sigma$ of the respective posterior distributions.}
    \label{fig:2C40}
\end{figure*}

\begin{figure*} [h]
    \centering
    \includegraphics[width=\linewidth]{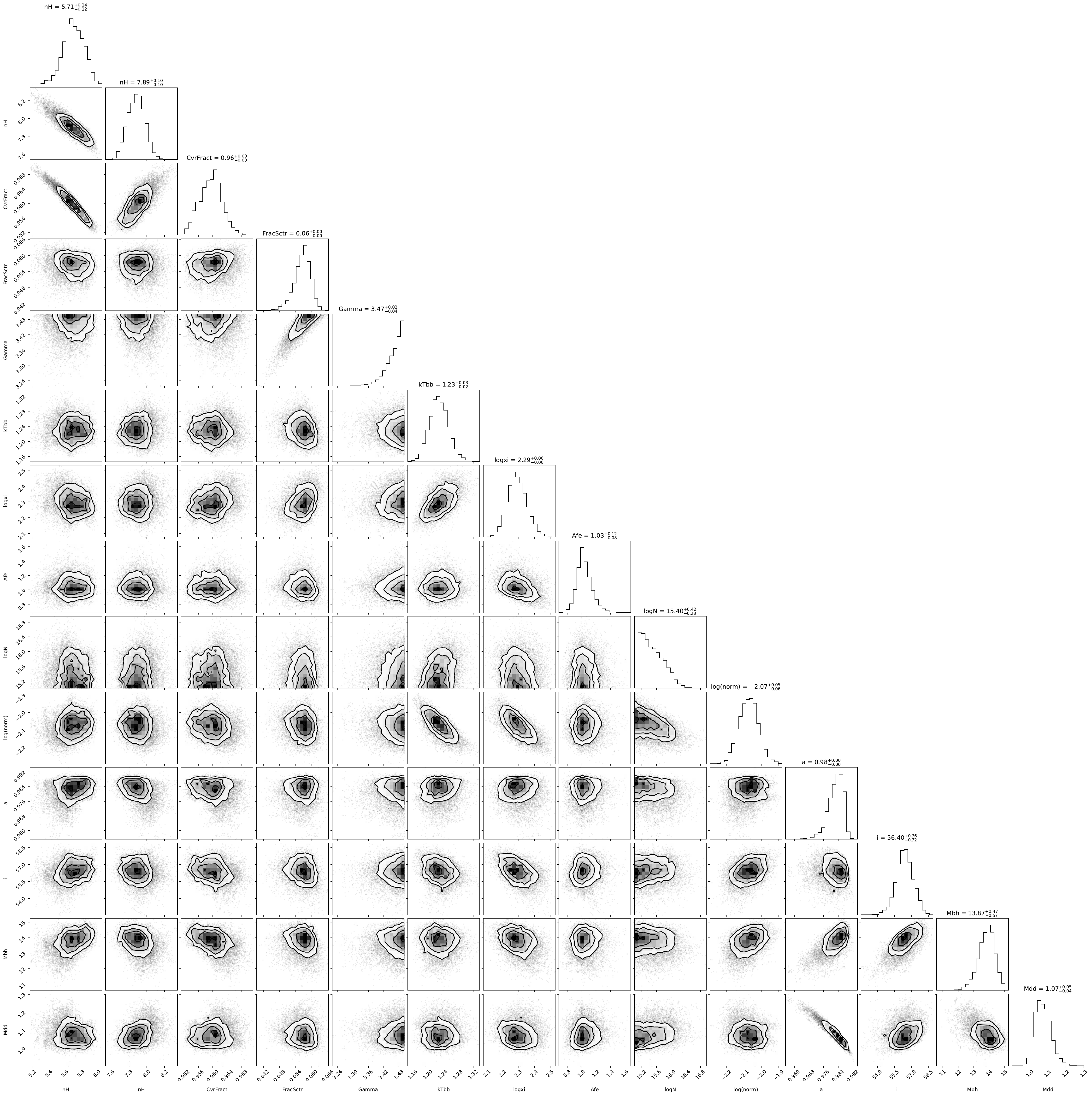}
    \caption{Corner plot of the Bayesian parameter estimation of Obs 2 using \texttt{Model 1}. The error values of each of the parameters correspond to $1\sigma$ of the respective posterior distributions.}
    \label{fig:2N40}
\end{figure*}

\begin{figure*} [h]
    \centering
    \includegraphics[width=\linewidth]{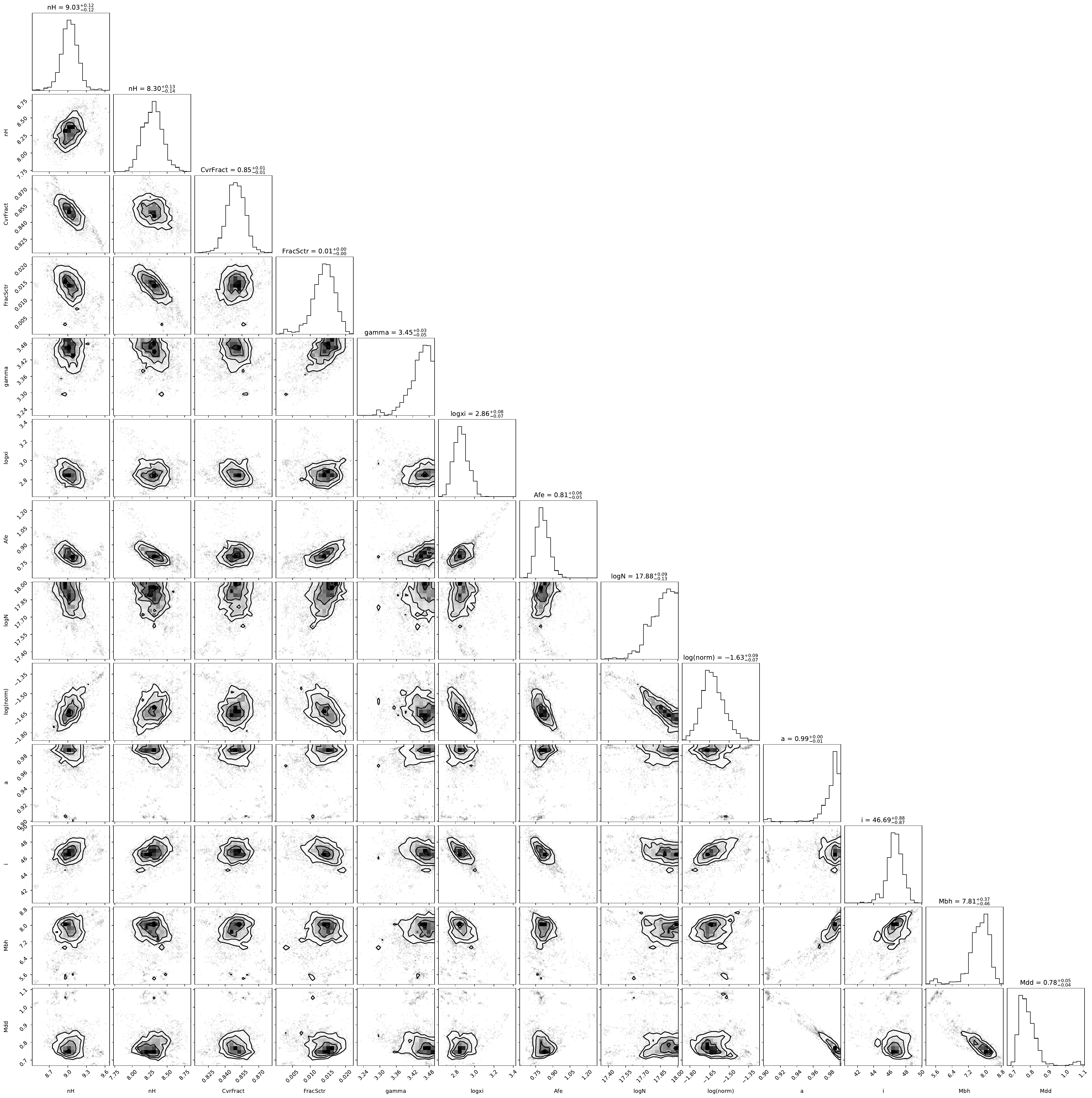}
    \caption{Corner plot of the Bayesian parameter estimation of Obs 3 using \texttt{Model 0}. The error values of each of the parameters correspond to $1\sigma$ of the respective posterior distributions.}
    \label{fig:3C40}
\end{figure*}

\begin{figure*} [h]
    \centering
    \includegraphics[width=\linewidth]{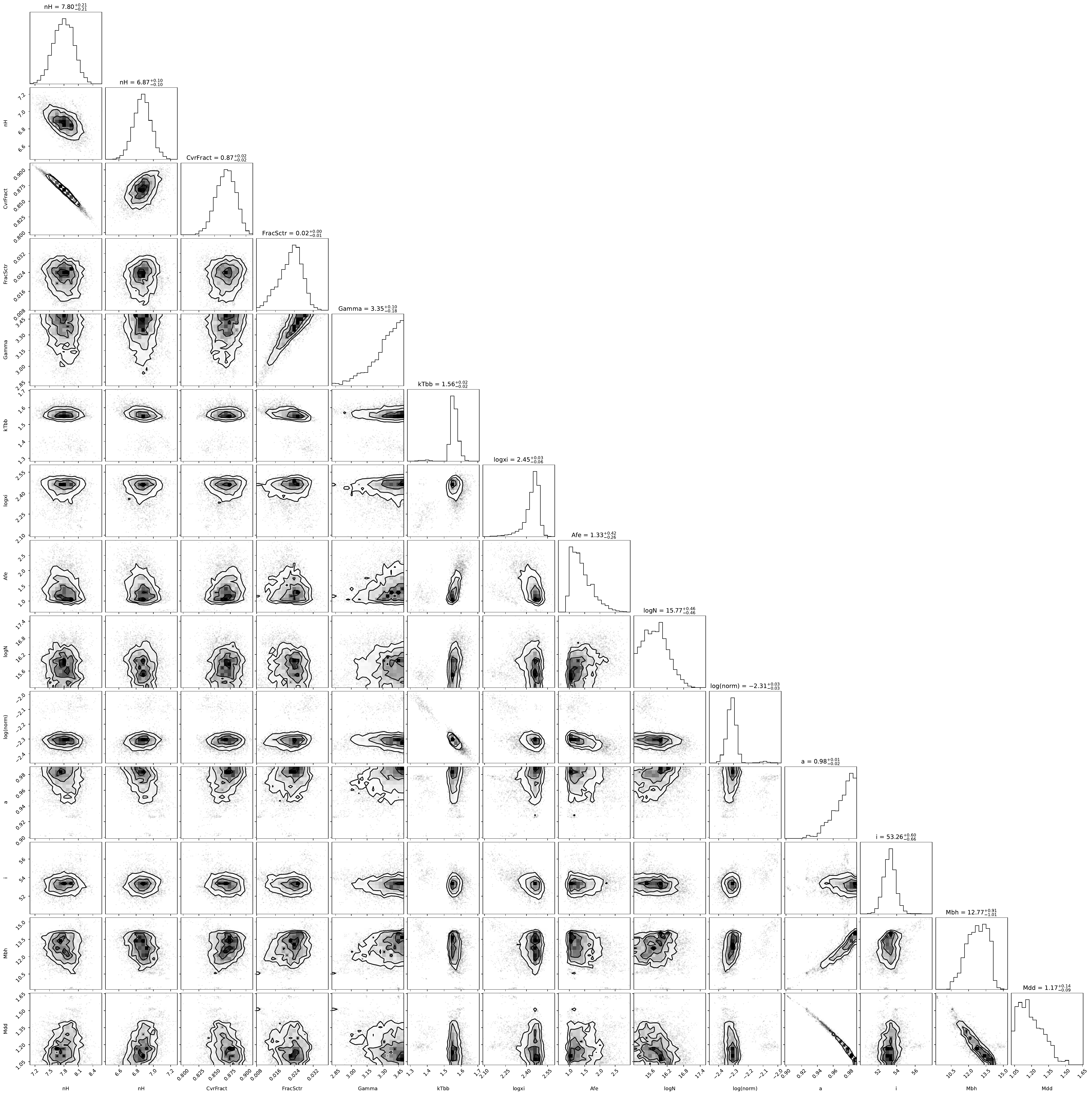}
    \caption{Corner plot of the Bayesian parameter estimation of Obs 3 using \texttt{Model 1}. The error values of each of the parameters correspond to $1\sigma$ of the respective posterior distributions.}
    \label{fig:3N40}
\end{figure*}

Tables \ref{priorCP} and \ref{priorNS} summarize the prior distributions adopted for the Bayesian analyses of \texttt{Model 0} (\texttt{relxillCp} configuration) and \texttt{Model 1} (\texttt{relxillCp} configuration), respectively. Tables \ref{bxaCP} and \ref{bxaNS} present the corresponding posterior parameter estimates derived from the Bayesian inference for the two models. Figures \ref{fig:1C40}-\ref{fig:3N40} show the corner plots of the posterior distributions obtained from the Bayesian analyses of the individual observations. These figures illustrate the marginalized posterior distributions of the fitted parameters together with their correlations. The parameter correlations and their physical implications are discussed in detail in Section~\ref{s-bxa}.

Finally, Tables \ref{qCP} and \ref{qNS} present the results of additional robustness tests performed by varying the emissivity index in the spectral fitting. These tables report the best-fit parameter values for Obs 1 obtained using the \texttt{relxillCp} and \texttt{relxillNS} models, respectively.

\begin{table*} [h]
	\centering
	\caption{Best-fit values with \texttt{relxillCp} model.}
	\label{CP}
	\renewcommand\arraystretch{1.25}
	\begin{tabular}{llllllll}
		\hline\hline
		Component & Parameter & Obs 1 && Obs 2 && Obs 3 &\\
		\hline
        & & frozen $M_{\rm BH}$ & free $M_{\rm BH}$ & frozen $M_{\rm BH}$ & free $M_{\rm BH}$ & frozen $M_{\rm BH}$ & free $M_{\rm BH}$ \\
		\hline
		
		\texttt{TBabs} & $N_{\rm H}$ (10$^{22}$ cm$^{-2}$)  & $7.61_{-0.52}^{+0.21}$ & $7.53_{-0.49}^{+0.44}$ & $8.19_{-0.45}^{+0.44}$ & $8.16_{-0.47}^{+0.46}$ & $9.2_{-0.37}^{+0.33}$ & $8.76_{-0.44}^{+0.39}$ \\

		\hline
		
		\texttt{pcfabs} & $N_{\rm H}$ (10$^{22}$ cm$^{-2}$) & $7.56_{-0.25}^{+0.23}$ & $7.53_{-0.31}^{+0.24}$ & $7.54_{-0.24}^{+0.24}$ & $8.07_{-0.27}^{+0.26}$ & $8.56_{-0.24}^{+0.27}$ &  $8.12_{-0.34}^{+0.29}$ \\

		& $CvrFract$ & $0.91_{-0.02}^{+0.02}$ & $0.91_{-0.02}^{+0.02}$ & $0.88_{-0.029}^{+0.025}$  & $0.9_{-0.025}^{+0.021}$ & $0.85_{-0.019}^{+0.023}$ & $0.88_{-0.026}^{+0.026}$ \\
		\hline
		
		\texttt{xstar} & $column$ ($10^{23}$)& $1.28_{-0.13}^{+0.17}$ & $1.28_{-0.13}^{+0.18}$ & $1.27_{-0.12}^{+0.18}$ & $1.3_{-0.13}^{+0.2}$ & $1.41_{-0.26}^{+0.13}$ &  $1.32_{-0.82}^{+0.16}$\\ 
		
		& $rlog\xi$ & $4.1_{-0.075}^{+0.126}$ & $4.1_{-0.075}^{+0.126}$ & $3.99_{-0.065}^{+0.126}$  & $4_{-0.072}^{+0.138}$ & $4.05_{-0.094}^{+0.087}$ & $3.94_{-0.047}^{+0.114}$\\
		\hline
		
		\texttt{Gaussian} & $LineE$ (keV)& $1.72_{-0.012}^{+0.013}$ & $1.72_{-0.013}^{+0.013}$ & $1.72_{-0.012}^{+0.013}$ & $1.73_{-0.012}^{+0.012}$ & $1.72_{-0.017}^{+0.017}$ &  $1.73_{-0.015}^{+0.016}$\\
		
		& $norm$ & $0.105_{-0.02}^{+0.02}$  & $0.104_{-0.02}^{+0.02}$ & $0.099_{-0.02}^{+0.02}$ &   $0.12_{-0.023}^{+0.024}$ & $0.125_{-0.031}^{+0.032}$  & $0.131_{-0.03}^{+0.03}$\\
		\hline
		
		\texttt{relxillCp} & Index$_1$ & 3* & 3* & 3* & 3* & 3* & 3*\\
		
		& Index$_2$ & 3* & 3* & 3* & 3* & 3* & 3*\\
		
		& $r_{in}$ ($r_g$) & -1* & -1* & -1* & -1* & -1* & -1*\\
		
		& $r_{br}$ ($r_g$) & 15* & 15* & 15* & 15* & 15* & 15*\\
		
		& $r_{out}$ ($r_g$) & 400* & 400* & 400* & 400* & 400* & 400*\\
		
		& $\Gamma$ & $2.99_{-0.13}^{+0.33}$ & $2.99_{-0.12}^{+0.11}$ & $3.42_{-0.14}^{p}$ & $3.5_{-0.16}^{p}$ & $3.5_{-0.11}^{p}$ & $3.5_{-0.16}^{p}$\\
		
		& $log(\xi)$ & $3.26_{-0.17}^{+0.16}$ & $3.29_{-0.36}^{+0.14}$ & $2.01_{-0.13}^{+0.06}$ &  $2.71_{-0.11}^{+0.04}$ & $2.68_{-0.11}^{+0.04}$ & $2.76_{-0.13}^{+0.1}$\\
		
		& $A_{Fe}$ (solar) & $1.6_{-0.18}^{+1.8}$ & $1.77_{-0.66}^{+3.88}$ & $0.96_{-0.08}^{+0.22}$  & $0.84_{-0.04}^{+0.15}$ & $0.62_{-0.05}^{+0.04}$ &  $0.8_{-0.05}^{+0.11}$\\
		
		& $log(n_{\rm e})$ (cm$^{-3}$) & $17.92_{-0.34}^{+0.17}$ &  $18.0_{-0.77}^{+0.08}$ & $18.3_{-1.01}^{+0.65}$ & $17.98_{-0.64}^{+0.04}$ & $17.98_{-0.36}^{+0.02}$ & $17.98_{-0.5}^{+0.03}$\\
		
		& $norm$ & $0.015_{-0.014}^{+0.026}$ & $0.013_{-0.007}^{+0.013}$ & $0.021_{-0.011}^{+0.019}$ & $0.022_{-0.009}^{+0.007}$ & $0.029_{-0.003}^{+0.008}$ & $0.025_{-0.006}^{+0.022}$\\
		\hline
		
		\texttt{simplcutx} & $f_{\rm sc}$ & $0.017_{p}^{+0.04}$ & $0.018_{p}^{+0.042}$ & $0.007_{-0.006}^{+0.057}$ & $0.038_{-0.015}^{+0.002}$ & $0.014_{-0.003}^{+0.002}$ & $0.014_{-0.003}^{+0.002}$\\   
		
		& $kT_e$ (keV) & 100* & 100* & 100* & 100* & 100* & 100* \\
		\hline
		
		\texttt{kerrbb} & $a_*$ & $0.983_{-0.009}^{p}$ &  $0.984_{-0.052}^{+0.006}$ & $0.99_{-0.009}^{p}$ & $0.99_{-0.01}^{p}$ & $0.99_{-0.01}^{p}$ & $0.99_{-0.091}^{p}$ \\
		
		& $i$ (degrees) & $51.3_{-3.17}^{2.2}$ & $50.79_{-2.68}^{+2.82}$ & $49.79_{-0.53}^{+1.16}$ & $50.09_{-1.87}^{+1.89}$ & $52.21_{-0.42}^{+0.26}$ & $47.14_{-4.77}^{+0.82}$ \\
		
		& $M_{\rm BH}$ ($\mathrm{M}_\odot$) & $10^*$ & $9.96_{-2.81}^{+0.44}$ & $10^*$ & $9.27_{-1.19}^{+0.67}$ & $10^*$ & $8.06_{-1.51}^{+0.36}$ \\
		
		& $\dot{M}_{\rm BH}$ & $1.02_{-0.16}^{+0.14}$ & $1.01_{-0.01}^{+0.11}$ & $0.87_{-0.02}^{+0.08}$ &  $0.8_{-0.03}^{+0.16}$ & $0.73_{-0.03}^{+0.02}$ & $0.71_{-0.03}^{+0.06}$\\
		
		& $d_{\rm BH}$ (kpc) & $10^*$ & $10^*$ & $10^*$ & $10^*$ & $10^*$ & $10^*$\\
		\hline
		
		\texttt{mbpo} & $d\Gamma_1$ & $0.197_{-0.025}^{+0.014}$ & $0.199_{-0.025}^{+0.023}$ & $0.204_{-0.022}^{+0.02}$ & $0.212_{-0.012}^{+0.02}$ & $0.195_{-0.019}^{+0.019}$ &  $0.219_{-0.02}^{+0.021}$ \\
		
		\textit{NICER} & $d\Gamma_2$ & $0.068_{-0.092}^{+0.048}$ & $0.067_{-0.091}^{+0.049}$ & $0.087_{-0.283}^{+0.086}$ & $0.079_{-0.074}^{+0.085}$ & $0.106_{-0.142}^{+0.078}$ & $0.102_{-0.082}^{+0.071}$ \\
		
		& $E_{br}$ & $6.04_{-0.62}^{+0.8}$ & $6.02_{-0.66}^{+0.8}$ & $6.63_{-0.84}^{+1.39}$ &  $6.65_{-0.97}^{+1.3}$ & $6.72_{-1.85}^{+1.39}$ & $6.58_{-1.53}^{+0.93}$ \\
		
		& $N$ & $0.989_{-0.019}^{+0.02}$ & $0.989_{-0.02}^{+0.021}$ & $1.017_{-0.027}^{+0.032}$ & $1.019_{-0.032}^{+0.031}$ & $1.022_{-0.071}^{+0.02}$ & $1.02_{-0.054}^{+0.024}$\\
		
		\hline
		\texttt{mbpo} \\
		\textit{NuSTAR} B& $N$ & $0.987_{-0.002}^{+0.002}$ & $0.987_{-0.002}^{+0.002}$ & $0.992_{-0.002}^{+0.002}$ & $0.992_{-0.002}^{+0.002}$ & $0.989_{-0.002}^{+0.002}$ &  $0.989_{-0.002}^{+0.002}$\\
		\hline
		
		& $\chi^2$/d.o.f & $944.6/909$ & $943.5/908$& $945.0/855$ & $936.4/854$ & $910.1/842$ & $911.2/841$\\
		\hline\hline
	\end{tabular}
	\\
	\textit{Note.} Parameters with $^*$ are fixed during the fit and the symbol $p$ denotes parameter hits its upper or lower boundary. 
\end{table*}

\begin{table*} [h]
	\centering
	\caption{Best-fit values with \texttt{relxillNS} model.}
	\label{NS}
	\renewcommand\arraystretch{1.25}
	\begin{tabular}{llllllll}
		\hline\hline
		Component & Parameter & Obs 1 && Obs 2 && Obs 3 &\\
		\hline
        & & frozen $M_{\rm BH}$ & free $M_{\rm BH}$ & frozen $M_{\rm BH}$ & free $M_{\rm BH}$ & frozen $M_{\rm BH}$ & free $M_{\rm BH}$ \\
		\hline
		
		\texttt{TBabs} & $N_{\rm H}$ (10$^{22}$ cm$^{-2}$)& $5.97_{-0.68}^{+0.69}$ & $6.06_{-0.37}^{+0.78}$ & $4.21_{-0.27}^{+0.28}$ & $3.72_{-0.75}^{+0.9}$ & $7.63_{-0.57}^{+0.52}$ & $7.8_{-0.53}^{+0.44}$\\
		
		\hline
		
		\texttt{pcfabs} & $N_{\rm H}$ (10$^{22}$ cm$^{-2}$) & $7.34_{-0.41}^{+0.39}$ & $7.29_{-0.39}^{+0.42}$  & $8.61_{-0.35}^{+0.35}$ & $8.32_{-0.71}^{+0.59}$ & $6.99_{-0.26}^{+0.24}$ & $6.89_{-0.26}^{+0.22}$ \\

		& $CvrFract$ & $0.957_{-0.024}^{+0.015}$ & $0.954_{-0.031}^{+0.013}$ & $0.98_{-0.091}^{+0.005}$ & $0.99_{-0.085}^{+0.004}$ & $0.88_{-0.054}^{+0.023}$ & $0.864_{-0.04}^{+0.037}$ \\
		\hline
		
		\texttt{xstar} & $column$ ($10^{23}$)& $1.23_{-0.15}^{+0.15}$ & $1.25_{-0.14}^{+0.15}$ & $1.1_{-0.36}^{+0.37}$ & $1.17_{-0.13}^{+0.19}$ & $1.22_{-0.77}^{+0.12}$ & $1.18_{-0.77}^{+0.14}$ \\ 
		
		& $rlog\xi$ & $4.11_{-0.09}^{+0.1}$ & $4.11_{-0.08}^{+0.14}$ & $3.95_{-0.03}^{+0.03}$ & $4_{-0.08}^{+0.12}$ & $3.93_{-0.04}^{+0.08}$ & $3.93_{-0.04}^{+0.09}$ \\
		\hline
		
		\texttt{Gaussian} & $LineE$ (keV)& $1.71_{-0.013}^{+0.013}$ & $1.71_{-0.013}^{+0.013}$ & $1.71_{-0.01}^{+0.01}$ & $1.71_{-0.02}^{+0.01}$ & $1.71_{-0.017}^{+0.016}$ & $1.71_{-0.016}^{+0.016}$ \\
		
		& $norm$ & $0.12_{-0.02}^{+0.02}$ & $0.12_{-0.02}^{+0.02}$ & $0.12_{-0.012}^{+0.012}$ & $0.1_{-0.015}^{+0.014}$ & $0.12_{-0.02}^{+0.02}$ & $0.11_{-0.02}^{+0.02}$ \\
		\hline
		
		\texttt{relxillNS} & Index$_1$ & 3* & 3* & 3* & 3* & 3* & 3*\\
		
		& Index$_2$ & 3* & 3* & 3* & 3* & 3* & 3*\\
		
		& $r_{in}$ ($r_g$) & -1* & -1* & -1* & -1* & -1* & -1*\\
		
		& $r_{br}$ ($r_g$) & 15* & 15* & 15* & 15* & 15* & 15*\\
		
		& $r_{out}$ ($r_g$) & 400* & 400* & 400* & 400* & 400* & 400*\\
		
		& $kT_{bb}$ (keV)& $1.11_{-0.032}^{+0.126}$ & $1.06_{-0.014}^{+0.07}$ & $0.8_{-0.012}^{+0.025}$ & $0.98_{-0.036}^{+0.032}$ & $1.27_{-0.13}^{+0.11}$ & $1.61_{-0.12}^{+0.07}$ \\
		
		& $log(\xi)$ & $2.48_{-0.17}^{+0.26}$ & $2.47_{-0.08}^{+0.22}$ & $2.37_{-0.02}^{+0.03}$ & $2.76_{-0.19}^{+0.06}$ & $2.25_{-0.03}^{+0.21}$ & $2.12_{-0.1}^{+0.12}$ \\
		
		& $A_{Fe}$ (solar) & $0.97_{p}^{+1.09}$ & $0.99_{p}^{+0.46}$ & $3.55_{-0.14}^{+0.13}$ & $3.2_{-0.29}^{+0.14}$ & $1.0_{-0.15}^{+0.2}$ & $1.0_{-0.1}^{+0.15}$ \\
		
		& $log(n_{\rm e})$ (cm$^{-3}$) & $17.11_{p}^{+1.32}$ & $17.11_{p}^{+0.69}$ & $15.0_{p}^{+0.98}$ & $18.0_{-0.51}^{+0.27}$ & $15_{p}^{+0.4}$ & $17.99_{-0.72}^{+0.22}$ \\
		
		& $norm$ & $0.007_{-0.002}^{+0.002}$ & $0.007_{-0.002}^{+0.002}$ & $0.039_{-0.001}^{+0.001}$ & $0.015_{-0.003}^{+0.005}$ & $0.01_{-0.002}^{+0.003}$ & $0.007_{-0.001}^{+0.002}$\\
		\hline
		
		\texttt{simplcutx} & $\Gamma$ & $3.42_{-0.33}^{p}$ & $3.47_{-0.35}^{p}$ & $3.5_{-0.22}^{p}$ & $3.39_{-0.35}^{p}$ & $3.5_{-0.18}^{p}$ & $2.84_{-0.56}^{p}$\\
		
		& $f_{\rm sc}$ & $0.067_{-0.024}^{+0.008}$  & $0.074_{-0.023}^{+0.002}$ & $0.065_{-0.016}^{+0.016}$ & $0.052_{-0.017}^{+0.03}$ & $0.036_{-0.008}^{+0.002}$ & $0.03_{-0.006}^{+0.007}$ \\ 
		
		& $kT_e$ (keV) & 100* & 100* & 100* & 100* & 100* & 100* \\
		\hline
		
		\texttt{kerrbb} & $a_*$ & $0.928_{-0.022}^{+0.022}$ & $0.845_{-0.171}^{+0.084}$ & $0.949_{-0.014}^{+0.015}$ & $0.908_{-0.065}^{+0.031}$ & $0.903_{-0.021}^{+0.005}$ & $0.953_{-0.016}^{+0.014}$\\
		
		& $i$ (degrees) & $56.95_{-1.75}^{+1.12}$ & $56.58_{-2.09}^{+2.04}$ & $57.57_{-0.44}^{+0.45}$ & $50.07_{-1.65}^{+1.48}$ & $53.83_{-1.2}^{+0.39}$ & $57.59_{-0.52}^{+0.47}$\\
		
		& $M_{\rm BH}$ ($\mathrm{M}_\odot$) & $10^*$ & $7.7_{-1.0}^{+1.2}$ & $10^*$ & $8.96_{-0.33}^{+0.57}$  & $10^*$ & $12.59_{-2.95}^{+2.7}$\\
		
		& $\dot{M}_{\rm BH}$ & $1.45_{-0.19}^{+0.16}$ & $1.82_{-0.41}^{+0.11}$ & $1.14_{-0.17}^{+0.2}$ & $0.98_{-0.07}^{+0.09}$ & $1.58_{-0.13}^{+0.08}$ & $1.41_{-0.14}^{+0.03}$ \\
		
		& $d_{\rm BH}$ (kpc) & $10^*$ & $10^*$ & $10^*$ & $10^*$ & $10^*$ & $10^*$\\
		\hline
		
		\texttt{mbpo} & $d\Gamma_1$ & $0.209_{-0.024}^{+0.03}$ & $0.205_{-0.027}^{+0.03}$ & $0.21_{-0.005}^{+0.005}$ & $0.213_{-0.021}^{+0.018}$ & $0.22_{-0.02}^{+0.02}$ & $0.205_{-0.017}^{+0.02}$\\
		
		\textit{NICER} &$d\Gamma_2$ & $0.066_{-0.079}^{+0.057}$ & $0.065_{-0.085}^{+0.051}$ & $0.087_{-0.059}^{+0.058}$ & $0.078_{-0.076}^{+0.073}$ & $0.108_{-0.08}^{+0.085}$ & $0.125_{-0.081}^{+0.063}$\\
		
		&$E_{br}$ & $5.96_{-0.91}^{+0.76}$ & $6.0_{-0.8}^{+0.78}$ & $6.63_{-0.1}^{+0.1}$ & $6.62_{-0.49}^{+1.24}$ & $6.55_{-1.46}^{+0.55}$ & $6.55_{-1.01}^{+1.03}$ \\
		
		& $N$ & $0.989_{-0.029}^{+0.015}$ & $0.989_{-0.026}^{+0.02}$ & $1.019_{-0.03}^{0.03}$ & $1.02_{-0.03}^{+0.037}$ & $1.022_{-0.055}^{+0.014}$ & $1.019_{-0.043}^{+0.028}$\\
		
		\hline
		\texttt{mbpo} \\
		\textit{NuSTAR} B& $N$ & $0.987_{-0.002}^{+0.002}$ & $0.987_{-0.002}^{+0.002}$ & $0.992_{-0.002}^{+0.002}$ & $0.992_{-0.002}^{+0.002}$ & $0.989_{-0.002}^{+0.002}$ & $0.989_{-0.002}^{+0.002}$\\
		\hline

		& $\chi^2$/d.o.f & $938.2/908$ & $937.2/907$ & $914/854$ & $895/853$& $859/841$ & $852.7/840$ \\
		\hline\hline
	\end{tabular}
	\\
	\textit{Note.} Parameters with $^*$ are fixed during the fit and the symbol $p$ denotes parameter hits its upper or lower boundary.
\end{table*}

\begin{table*} [h]
	\centering
	\caption{Best-fit values of the simultaneous fit with \texttt{relxillCp} model.}
	\label{CP_all}
	\renewcommand\arraystretch{1.25}
	\begin{tabular}{lllll}
		\hline\hline
		Component & Parameter & Obs 1 & Obs 2 & Obs 3 \\
		
		\hline
		
		\texttt{TBabs} & $N_{\rm H}$ (10$^{22}$ cm$^{-2}$)  & $8.2_{-0.2}^{+0.19}$ \\

		\hline
		
		\texttt{pcfabs} & $N_{\rm H}$ (10$^{22}$ cm$^{-2}$) & $7.63_{-0.12}^{+0.11}$ & $7.65_{-0.23}^{+0.16}$ & $6.86_{-0.09}^{+0.14}$\\

		& $CvrFract$ & $0.88_{-0.012}^{+0.013}$ & $0.87_{-0.023}^{+0.015}$ & $0.87_{-0.016}^{+0.017}$  \\
		\hline
		
		\texttt{xstar} & $column$ ($10^{23}$)& $1.1_{-0.09}^{+0.12}$ & $1.36_{-0.11}^{+0.13}$ & $1.49_{-0.22}^{+0.22}$ \\ 
		
		& $rlog\xi$ & $3.98_{-0.06}^{+0.11}$ & $4.05_{-0.08}^{+0.08}$ & $4.09_{-0.04}^{+0.07}$ \\
		\hline
		
		\texttt{Gaussian} & $LineE$ (keV)& $1.73_{-0.005}^{+0.007}$ \\
		
		& $norm$ & $0.092_{-0.013}^{+0.015}$  & $0.105_{-0.015}^{+0.016}$ & $0.099_{-0.012}^{+0.016}$\\
		\hline
		
		\texttt{relxillCp} & Index$_1$ & 3* & 3* & 3* \\
		
		& Index$_2$ & 3* & 3* & 3*\\
		
		& $r_{in}$ ($r_g$) & -1* & -1* & -1* \\
		
		& $r_{br}$ ($r_g$) & 15* & 15* & 15* \\
		
		& $r_{out}$ ($r_g$) & 400* & 400* & 400* \\
		
		& $\Gamma$ & $3.14_{-0.04}^{+0.07}$ & $2.69_{-0.1}^{+0.24}$ & $2.58_{-0.08}^{+0.14}$ \\
		
		& $log(\xi)$ & $2.2_{-0.07}^{+0.08}$ & $3.3_{-0.11}^{+0.1}$ & $2.7_{-0.05}^{+0.08}$ \\
		
		& $A_{Fe}$ (solar) & $1.09_{-0.12}^{+0.16}$  \\
		
		& $log(n_{\rm e})$ (cm$^{-3}$) & $18.35_{-0.16}^{+0.13}$ & $18.03_{-0.62}^{+0.22}$ & $18.69_{-0.19}^{+0.16}$ \\
		
		& $norm$ & $0.022_{-0.005}^{+0.006}$ & $0.02_{-0.005}^{+0.005}$ & $0.015_{-0.003}^{+0.001}$ \\
		\hline
		
		\texttt{simplcutx} & $f_{\rm sc}$ & $0.007_{-0.002}^{+0.003}$ & $0.003_{-0.002}^{+0.005}$ & $0.0_{p}^{+0.003}$ \\
		
		& $kT_e$ (keV) & 100* & 100* & 100* \\
		\hline
		
		\texttt{kerrbb} & $a_*$ & $0.967_{-0.008}^{+0.004}$ \\
		
		& $i$ (degrees) & $50.71_{-0.68}^{+0.85}$  \\
		
		& $M_{\rm BH}$ ($\mathrm{M}_\odot$) & $9.14_{-0.39}^{+0.91}$ \\
		
		& $\dot{M}_{\rm BH}$ & $1.11_{-0.02}^{+0.05}$ & $1.08_{-0.04}^{+0.08}$ & $1.07_{-0.04}^{+0.09}$\\
		
		& $d_{\rm BH}$ (kpc) & $10^*$\\
		\hline
		
		\texttt{mbpo} & $d\Gamma_1$ & $0.177_{-0.011}^{+0.01}$  \\
		
		\textit{NICER} & $d\Gamma_2$ & $0.026_{-0.044}^{+0.043}$ \\
		
		& $E_{br}$ & $6.59_{-0.29}^{+0.29}$ \\
		
		& $N$ & $1.003_{-0.008}^{+0.007}$ \\
		
		\hline
		\texttt{mbpo} \\
		\textit{NuSTAR} B& $N$ & $0.989_{-0.001}^{+0.001}$ \\
		\hline
		
		& $\chi^2$/d.o.f & $3075.1/2627$ \\
		\hline\hline
	\end{tabular}
	\\
	\textit{Note.} The symbol $p$ denotes parameter hits its upper or lower boundary.
\end{table*}

\begin{table*} [h]
	\centering
	\caption{Best-fit values of the simultaneous fit with \texttt{relxillNS} model.}
	\label{NS_all}
	\renewcommand\arraystretch{1.25}
	\begin{tabular}{lllll}
		\hline\hline
		Component & Parameter & Obs 1 & Obs 2 & Obs 3 \\
		
		\hline
		
		\texttt{TBabs} & $N_{\rm H}$ (10$^{22}$ cm$^{-2}$)  & $7.19_{-0.13}^{+0.26}$ \\

		\hline
		
		\texttt{pcfabs} & $N_{\rm H}$ (10$^{22}$ cm$^{-2}$) & $7.1_{-0.15}^{+0.12}$ & $7.36_{-0.1}^{+0.12}$ & $7_{-0.15}^{+0.12}$\\

		& $CvrFract$ & $0.9_{-0.012}^{+0.01}$ & $0.91_{-0.018}^{+0.009}$ & $0.9_{-0.02}^{+0.01}$  \\
		\hline
		
		\texttt{xstar} & $column$ ($10^{23}$)& $1.24_{-0.15}^{+0.13}$ & $1.19_{-0.1}^{+0.21}$ & $1.19_{-0.77}^{+0.13}$ \\ 
		
		& $rlog\xi$ & $4.1_{-0.09}^{+0.08}$ & $3.97_{-0.06}^{+0.11}$ & $3.94_{-0.04}^{+0.1}$ \\
		\hline
		
		\texttt{Gaussian} & $LineE$ (keV)& $1.73_{-0.007}^{+0.007}$ \\
		
		& $norm$ & $0.081_{-0.014}^{+0.014}$  & $0.096_{-0.016}^{+0.016}$ & $0.086_{-0.015}^{+0.015}$\\
		\hline
		
		\texttt{relxillNS} & Index$_1$ & 3* & 3* & 3*\\
		
		& Index$_2$ & 3* & 3* & 3*\\
		
		& $r_{in}$ ($r_g$) & -1* & -1* & -1*\\
		
		& $r_{br}$ ($r_g$) & 15* & 15* & 15* \\
		
		& $r_{out}$ ($r_g$) & 400* & 400* & 400* \\
		
		& $kT_{bb}$ (keV) & $1.29_{-0.226}^{+0.17}$ & $1.69_{-0.051}^{+0.033}$ & $1.67_{-0.058}^{+0.03}$ \\
		
		& $log(\xi)$ & $2.25_{-0.12}^{+0.37}$ & $1.98_{-0.38}^{+0.48}$ & $1.02_{p}^{+0.23}$ \\
		
		& $A_{Fe}$ (solar) & $2.68_{-0.14}^{+0.19}$  \\
		
		& $log(n_{\rm e})$ (cm$^{-3}$) & $18.76_{-1.67}^{+0.34}$ & $18.46_{-0.12}^{+0.4}$ & $17.98_{-0.33}^{+0.09}$ \\
		
		& $norm$ & $0.004_{-0.001}^{+0.001}$ & $0.003_{-0.001}^{+0.001}$ & $0.004_{-0.001}^{+0.001}$ \\
		\hline
		
		\texttt{simplcutx} & $\Gamma$ & $3.32_{-0.21}^{p}$ & $2.01_{-0.42}^{+0.79}$ & $1.94_{-0.56}^{+1.64}$ \\
		
		& $f_{\rm sc}$ & $0.054_{-0.012}^{+0.015}$ & $0.003_{-0.001}^{+0.008}$ & $0.002_{-0.001}^{+0.001}$ \\
		
		& $kT_e$ (keV) & 100* & 100* & 100* \\
		\hline
		
		\texttt{kerrbb} & $a_*$ & $0.964_{-0.003}^{+0.006}$ \\
		
		& $i$ (degrees) & $56.92_{-0.52}^{+0.74}$  \\
		
		& $M_{\rm BH}$ ($\mathrm{M}_\odot$) & $12.68_{-0.86}^{+0.82}$ \\
		
		& $\dot{M}_{\rm BH}$ & $1.33_{-0.02}^{+0.02}$ & $1.29_{-0.03}^{+0.06}$ & $1.29_{-0.02}^{+0.01}$\\
		
		& $d_{\rm BH}$ (kpc) & $10^*$\\
		\hline
		
		\texttt{mbpo} & $d\Gamma_1$ & $0.201_{-0.01}^{+0.01}$  \\
		
		\textit{NICER} & $d\Gamma_2$ & $0.072_{-0.04}^{+0.032}$ \\
		
		& $E_{br}$ & $6.56_{-0.46}^{+0.22}$ \\
		
		& $N$ & $1.012_{-0.014}^{+0.007}$ \\
		
		\hline
		\texttt{mbpo} \\
		\textit{NuSTAR} B& $N$ & $0.989_{-0.001}^{+0.001}$ \\
		\hline
		
		& $\chi^2$/d.o.f & $2768.4/2626$ &&\\
		\hline\hline
	\end{tabular}
	\\
	\textit{Note.} The symbol $p$ denotes parameter hits its upper or lower boundary.
\end{table*}


\begin{table*} [h]
	\centering
	\caption{Prior distributions for Bayesian parameter estimation with \texttt{Model 0}} 
	\label{priorCP}
	\renewcommand\arraystretch{1.25}
	\begin{tabular}{llllllll}
		\hline\hline
		Component & Parameter & Prior type& Obs 1 & Obs 2 & Obs 3  \\
		\hline
		
		\texttt{TBabs} &$N_{\rm H}$ ($10^{22}$ cm$^{-2}$) & Gaussian & $7.6_{-3.0}^{+3.0}$ & $8.2_{-3.0}^{+3.0}$ & $9.2_{-3.0}^{+3.0}$ \\
		
		\hline
		
		\texttt{pcfabs} & $N_{\rm H}$ ($10^{22}$ cm$^{-2}$) & Gaussian & $7.5_{-1.5}^{+1.5}$ & $7.5_{-1.5}^{+1.5}$ & $8.5_{-1.5}^{+1.5}$ \\

		& $CvrFract$ & Gaussian & $0.87_{-0.12}^{+0.12}$ & $0.87_{-0.12}^{+0.12}$ & $0.87_{-0.12}^{+0.12}$\\
		\hline
		
		\texttt{relxillCp} & $\Gamma$ & Uniform & [2.0-3.5] & [2.0-3.5] & [2.0-3.5]\\
		
		& $log(\xi)$ & Uniform & [1.0-4.0] & [1.0-4.0] & [1.0-4.0]\\
		
		& $A_{Fe}$ (solar) & Uniform & [0.5-10.0] & [0.5-10.0] & [0.5-10.0]\\
		
		& $log(n_{\rm e})$ (cm$^{-3}$) & Uniform & [15.0-19.0] & [15.0-19.0] & [15.0-19.0]\\
		
		& $norm$ & Jeffreys & [$10^{-4}$-0.1] & [$10^{-4}$-0.1] & [$10^{-4}$-0.1]\\
		\hline
		
		\texttt{simplcutx} & $f_{\rm sc}$ & Uniform & [0.0-0.4] & [0.0-0.4] & [0.0-0.4]\\

		\hline
		
		\texttt{kerrbb} & $a_*$ & Uniform & [0.0-0.998] & [0.0-0.998]& [0.0-0.998]\\
		
		& $i$ (degrees)& Uniform & [30.0-85.0]& [30.0-85.0]& [30.0-85.0]\\
		
		& $M_{\rm BH}$ ($\mathrm{M}_\odot$)& Uniform & [5.0-20.0]& [5.0-20.0]& [5.0-20.0] \\
		
		& $\dot{M}_{\rm BH}$ & Gaussian & $1.05_{-1.0}^{+1.0}$ & $0.9_{-0.8}^{+0.8}$ & $0.9_{-0.8}^{+0.8}$\\
		
		
		\hline\hline
	\end{tabular}
\end{table*}

\begin{table*} [h]
	\centering
	\caption{Prior distributions for Bayesian parameter estimation with \texttt{Model 1}} 
	\label{priorNS}
	\renewcommand\arraystretch{1.25}
	\begin{tabular}{llllllll}
		\hline\hline
		Component & Parameter & Prior type& Obs 1 & Obs 2 & Obs 3  \\
		\hline
		
		\texttt{TBabs} &$N_{\rm H}$ ($10^{22}$ cm$^{-2}$) & Gaussian & $6.0_{-4.5}^{+4.5}$ & $4.5_{-2.0}^{+2.0}$ & $7.5_{-4.5}^{+4.5}$  \\
		
		\hline
		
		\texttt{pcfabs} & $N_{\rm H}$ ($10^{22}$ cm$^{-2}$) & Gaussian & $7.3_{-3.0}^{+3.0}$ & $8.6_{-3.0}^{+3.0}$ & $7.0_{-3.0}^{+3.0}$ \\

		& $CvrFract$ & Gaussian & $0.915_{-0.075}^{+0.075}$ & $0.945_{-0.045}^{+0.045}$ & $0.87_{-0.12}^{+0.12}$\\
		\hline
		
		\texttt{relxillNS} & $kT_{bb}$ (keV)& Uniform & [0.5-2.0] & [0.5-2.0] & [0.5-2.0]\\
		
		& $log(\xi)$ & Uniform & [1.0-4.0] & [1.0-4.0] & [1.0-4.0]\\
		
		& $A_{Fe}$ (solar) & Uniform & [0.5-10.0] & [0.5-10.0] & [0.5-10.0]\\
		
		& $log(n_{\rm e})$ (cm$^{-3}$) & Uniform & [15.0-19.0] & [15.0-19.0] & [15.0-19.0]\\
		
		& $norm$ & Jeffreys & [$10^{-5}$-0.1] & [$10^{-5}$-0.1] & [$10^{-5}$-0.1]\\
		\hline
		
		\texttt{simplcutx} & $f_{\rm sc}$ & Uniform & [0.0-0.4] & [0.0-0.4] & [0.0-0.4]\\
		
		&$\Gamma$ & Uniform & [2.0-3.5] & [2.0-3.5] & [2.0-3.5] \\
		
		\hline
		
		\texttt{kerrbb} & $a_*$ & Uniform & [0.0-0.998] & [0.0-0.998]& [0.0-0.998]\\
		
		& $i$ (degrees)& Uniform & [30.0-85.0]& [30.0-85.0]& [30.0-85.0]\\
		
		& $M_{\rm BH}$ ($\mathrm{M}_\odot$)& Uniform & [5.0-20.0]& [5.0-20.0]& [5.0-20.0] \\
		
		& $\dot{M}_{\rm BH}$ & Gaussian & $1.45_{-1.0}^{+1.0}$ & $1.15_{-0.8}^{+0.8}$ & $1.6_{-0.7}^{+0.7}$\\
		
		
		\hline\hline
	\end{tabular}
\end{table*}

\begin{table*} [h]
	\centering
	\caption{Bayesian parameter estimation with \texttt{Model 0}}
	\label{bxaCP}
	\renewcommand\arraystretch{1.75}
	\begin{tabular}{lllll}
		\hline\hline
		Component & Parameter &  Obs 1 & Obs 2 & Obs 3  \\
		&Speed&40&40&40\\
		\hline
		
		\texttt{TBabs} &$N_{\rm H}$ ($10^{22}$ cm$^{-2}$)& $7.41_{-0.28}^{+0.25}$ & $8.41_{-0.30}^{+0.28}$ & $9.03_{-0.20}^{+0.20}$\\
		
		\hline
		
		\texttt{pcfabs} & $N_{\rm H}$ ($10^{22}$ cm$^{-2}$)& $7.29_{-0.19}^{+0.15}$ & $7.85_{-0.20}^{+0.21}$ & $8.30_{-0.23}^{+0.22}$\\ 
		
		& $CvrFract$ & $0.91_{-0.01}^{+0.01}$ & $0.87_{-0.02}^{+0.02}$ & $0.85_{-0.01}^{+0.01}$\\
		\hline
		
		\texttt{relxillCp} & $\Gamma$ & $3.26_{-0.28}^{+0.24}$ & $3.43_{-0.08}^{+0.07}$ & $3.45_{-0.08}^{+0.05}$\\
		
		& $log(\xi)$ & $3.18_{-0.28}^{+0.51}$ & $2.94_{-0.17}^{+0.11}$ & $2.86_{-0.12}^{+0.14}$\\
		
		& $A_{Fe}$ (solar) & $3.19_{-0.89}^{+1.90}$ & $1.15_{-0.21}^{+0.27}$ & $0.81_{-0.08}^{+0.10}$ \\
		
		& $log(n_{\rm e})$ (cm$^{-3}$) & $17.77_{-0.56}^{+0.34}$ & $17.78_{-0.26}^{+0.21}$ & $17.88_{-0.22}^{+0.14}$\\
		
		& $norm$ &$0.008_{-0.003}^{+0.004}$ & $0.016_{-0.005}^{+0.008}$ & $0.024_{-0.006}^{+0.009}$\\
		\hline
		
		\texttt{simplcutx} & $f_{\rm sc}$ & $0.04_{-0.02}^{+0.02}$ & $0.04_{-0.01}^{+0.01}$ & $0.01_{-0.008}^{+0.003}$\\ 
		
		\hline
		
		\texttt{kerrbb} & $a_*$ & $0.97_{-0.01}^{+0.01}$ & $0.96_{-0.02}^{+0.03}$ & $0.99_{-0.02}^{+0.01}$\\
		
		& $i$ (degrees)& $50.18_{-2.15}^{+2.64}$ & $47.58_{-1.63}^{+1.68}$ & $46.69_{-1.44}^{+1.45}$\\
		
		& $M_{\rm BH}$ ($\mathrm{M}_\odot$)& $9.53_{-0.71}^{+1.05}$ & $7.59_{-0.77}^{+0.99}$ & $7.81_{-0.76}^{+0.61}$ \\
		
		& $\dot{M}_{\rm BH}$ & $1.16_{-0.09}^{+0.09}$ & $1.05_{-0.13}^{+0.09}$ & $0.78_{-0.06}^{+0.09}$ \\
		
		\hline

		\hline
		
		& $\ln (Z)$ & -594.81 & -595.80 & -530.12 \\
		& $\sigma_{\rm \ln (Z)}$ & 0.82 & 0.72 & 0.48 \\
		\hline\hline
	\end{tabular}
\end{table*}

\begin{table*} [h]
	\centering
	\caption{Bayesian parameter estimation with \texttt{Model 1}}
	\label{bxaNS}
	\renewcommand\arraystretch{1.75}
	\begin{tabular}{lllll}
		\hline\hline
		Component & Parameter & Obs 1 & Obs 2 & Obs 3 \\
		&Speed&40&40&40\\
		\hline
		
		\texttt{TBabs} &$N_{\rm H}$ ($10^{22}$ cm$^{-2}$) &$6.15_{-0.36}^{+0.29}$ & $5.71_{-0.19}^{+0.22}$ & $7.80_{-0.35}^{+0.34}$\\
		
		\hline
		
		\texttt{pcfabs} & $N_{\rm H}$ ($10^{22}$ cm$^{-2}$)&  $7.29_{-0.19}^{+0.20}$ & $7.89_{-0.17}^{+0.16}$ & $6.87_{-0.17}^{+0.17}$ \\ 
		
		& $CvrFract$ & $0.95_{-0.01}^{+0.01}$ & $0.96_{-0.01}^{+0.01}$ & $0.87_{-0.03}^{+0.03}$\\
		\hline
		
		\texttt{relxillNS} & $kT_{bb}$ (keV)&$1.12_{-0.07}^{+0.06}$ & $1.23_{-0.04}^{+0.04}$ & $1.56_{-0.03}^{+0.04}$\\
		
		& $log(\xi)$ & $2.51_{-0.14}^{+0.17}$ & $2.29_{-0.09}^{+0.11}$ & $2.45_{-0.09}^{+0.05}$\\
		
		& $A_{Fe}$ (solar) & $1.33_{-0.54}^{+0.75}$ & $1.03_{-0.15}^{+0.21}$ & $1.33_{-0.42}^{+0.70}$\\
		
		& $log(n_{\rm e})$ (cm$^{-3}$) & $16.58_{-1.51}^{+1.21}$ & $15.40_{-0.40}^{+0.69}$  & $15.77_{-0.77}^{+0.75}$\\
		
		& $norm$ & $0.006_{-0.001}^{+0.002}$ & $0.008_{-0.002}^{+0.002}$ & $0.005_{-0.001}^{+0.001}$\\
		\hline
		
		\texttt{simplcutx} & $f_{\rm sc}$ & $0.07_{-0.02}^{+0.01}$ & $0.06_{-0.01}^{+0.01}$ & $0.02_{-0.01}^{+0.01}$\\  
		
		& $\Gamma$ & $3.40_{-0.20}^{+0.10}$ & $3.47_{-0.07}^{+0.03}$ & $3.35_{-0.29}^{+0.15}$ \\ 
		
		\hline
		
		\texttt{kerrbb} & $a_*$ & $0.92_{-0.05}^{+0.03}$ & $0.98_{-0.02}^{+0.01}$ & $0.98_{-0.03}^{+0.02}$\\
		
		& $i$ (degrees)& $57.54_{-1.80}^{+1.86}$ & $56.40_{-1.19}^{+1.27}$ & $53.27_{-1.11}^{+1.07}$\\
		
		& $M_{\rm BH}$ ($\mathrm{M}_\odot$)&  $9.92_{-1.77}^{+1.35}$ & $13.87_{-0.94}^{+0.78}$ & $12.77_{-1.66}^{+1.51}$\\
		
		& $\dot{M}_{\rm BH}$ & $1.56_{-0.17}^{+0.22}$ &  $1.07_{-0.07}^{+0.11}$ & $1.17_{-0.15}^{+0.23}$ \\
		
		\hline
		
		\multicolumn{2}{c}{Inner disk temperature (keV)} & $1.52_{-0.04}^{+0.18}$ & $1.07_{-0.09}^{+0.08}$ & $1.11_{-0.23}^{+0.17}$ \\ 
		\hline
		
		& $\ln (Z)$ & -582.52 & -581.56 & -516.29\\
		& $\sigma_{\rm \ln (Z)}$ & 0.49 & 0.71 & 0.86 \\
		\multicolumn{2}{c}{$\Delta \ln (Z) = \ln (Z)_{\texttt{Model 1}} - \ln (Z)_{\texttt{Model 0}}$} & 12.29 & 14.24 & 13.83 \\
        BF$_{\texttt{Model 1}/\texttt{Model 0}}$ &($10^6$)&  0.22 & 1.53 & 1.01 \\
		\hline\hline
	\end{tabular}
	
\end{table*}

\begin{table*} [h]
	\centering
	\caption{Best-fit values of Obs 1 with \texttt{relxillCp} and varying emissivity index.}
	\label{qCP}
	\renewcommand\arraystretch{1.25}
	\begin{tabular}{lllll}
		\hline\hline
		Component & Parameter & $q_1 = q_2 = 3$ & $q_1 = q_2 = 5$& $q_1 = q_2 = 7$ \\
		\hline
		
		\texttt{TBabs} & $N_{\rm H}$ (10$^{22}$ cm$^{-2}$)& $7.61_{-0.52}^{+0.21}$ & $6.86_{-0.7}^{+0.61}$ & $7.36_{-0.26}^{+0.45}$ \\
		
		\hline
		
		\texttt{pcfabs} & $N_{\rm H}$ (10$^{22}$ cm$^{-2}$) &  $7.56_{-0.25}^{+0.23}$ & $7.2_{-0.52}^{+0.42}$ & $7.26_{-0.27}^{+0.24}$ \\

		& $CvrFract$ & $0.91_{-0.02}^{+0.02}$  & $0.94_{-0.024}^{+0.017}$ & $0.91_{-0.027}^{+0.024}$ \\
		\hline
		
		\texttt{xstar} & $column$ ($10^{23}$)& $1.28_{-0.13}^{+0.17}$ & $1.26_{-0.15}^{+0.18}$ & $1.38_{-0.2}^{+0.16}$ \\ 
		
		& $rlog\xi$ & $4.1_{-0.075}^{+0.126}$ & $4.09_{-0.09}^{+0.125}$ & $4.14_{-0.125}^{+0.109}$\\
		\hline
		
		\texttt{Gaussian} & $LineE$ (keV)& $1.72_{-0.012}^{+0.013}$ & $1.72_{-0.012}^{+0.013}$ & $1.72_{-0.013}^{+0.013}$\\
		
		& $norm$ & $0.105_{-0.02}^{+0.02}$ & $0.107_{-0.024}^{+0.023}$ & $0.096_{-0.019}^{+0.019}$ \\
		\hline
		
		\texttt{relxillCp} & $q_1$ & $3^*$ & $5^*$ & $7^*$ \\

        & $q_2$ & $3^*$ & $5^*$ & $7^*$ \\

        & $\Gamma$ & $2.99_{-0.13}^{+0.33}$ & $3.05_{-0.49}^{+0.23}$ & $3.23_{-0.11}^{+0.06}$ \\
		
		& $log(\xi)$ & $3.26_{-0.17}^{+0.16}$ & $2.86_{-0.59}^{+0.41}$ & $2.31_{-0.21}^{+0.09}$ \\
		
		& $A_{Fe}$ (solar) & $1.6_{-0.18}^{+1.8}$ & $2.77_{-1.44}^{+5.51}$ & $1.26_{-0.07}^{+0.81}$ \\
		
		& $log(n_{\rm e})$ (cm$^{-3}$) & $17.92_{-0.34}^{+0.17}$ & $18.07_{-0.8}^{+0.2}$ & $18.75_{-0.66}^{p}$ \\
		
		& $norm$ & $0.015_{-0.014}^{+0.026}$  & $0.016_{-0.01}^{+0.018}$ & $0.01_{-0.001}^{+0.004}$ \\
		\hline
		
		\texttt{simplcutx} & $f_{\rm sc}$ & $0.017_{p}^{+0.04}$ & $0.019_{p}^{+0.023}$ & $0.005_{p}^{+0.021}$ \\ 
		
		\hline
		
		\texttt{kerrbb} & $a_*$ & $0.983_{-0.009}^{p}$ & $0.948_{-0.012}^{+0.015}$ & $0.911_{-0.005}^{+0.01}$\\
		
		& $i$ (degrees) & $51.3_{-3.17}^{2.2}$ & $56.36_{-1.54}^{+1.43}$ & $59.49_{-0.64}^{+0.3}$ \\

        & $M_{\rm BH}$ ($\mathrm{M}_\odot$) & $10^*$ & $10^*$ & $10^*$ \\
		
		& $\dot{M}_{\rm BH}$ & $1.02_{-0.16}^{+0.14}$ & $1.3_{-0.27}^{+0.27}$ & $1.58_{-0.04}^{+0.07}$ \\

        & $d_{\rm BH}$ (kpc) & $10^*$ & $10^*$ & $10^*$ \\
		
		\hline
		
		\texttt{mbpo} & $d\Gamma_1$ & $0.197_{-0.025}^{+0.014}$ & $0.234_{-0.022}^{+0.035}$ & $0.194_{-0.024}^{+0.026}$ \\
		
		\textit{NICER} &$d\Gamma_2$ & $0.068_{-0.092}^{+0.048}$ & $0.095_{-0.078}^{+0.038}$ & $0.069_{-0.097}^{+0.024}$ \\
		
		&$E_{br}$ & $6.04_{-0.62}^{+0.8}$ & $5.37_{-0.6}^{+1.07}$ & $6.05_{-0.61}^{+0.84}$ \\
		
		& $N$ & $0.989_{-0.019}^{+0.02}$ & $0.973_{-0.017}^{+0.349}$ & $0.989_{-0.018}^{+0.021}$\\
		
		\hline
		\texttt{mbpo} \\
		\textit{NuSTAR} B & $N$ & $0.987_{-0.002}^{+0.002}$ & $0.987_{-0.002}^{+0.002}$ & $0.987_{-0.002}^{+0.002}$\\
		\hline

		& $\chi^2$/d.o.f & $944.6/909$ & $944/909$ & $956.6/909$  \\
		\hline\hline
	\end{tabular}
	\\
	\textit{Note.} Parameters with the symbol $p$ denotes parameter hits its upper or lower boundary.
\end{table*}

\begin{table*} [h]
	\centering
	\caption{Best-fit values of Obs 1 with \texttt{relxillNS} and varying emissivity index.}
	\label{qNS}
	\renewcommand\arraystretch{1.25}
	\begin{tabular}{lllll}
		\hline\hline
		Component & Parameter & $q_1 = q_2 = 3$ & $q_1 = q_2 = 5$& $q_1 = q_2 = 7$ \\
		\hline
		
		\texttt{TBabs} & $N_{\rm H}$ (10$^{22}$ cm$^{-2}$)& $5.97_{-0.68}^{+0.69}$ & $6.26_{-0.75}^{+0.71}$ & $6.27_{-0.93}^{+0.6}$ \\
		
		\hline
		
		\texttt{pcfabs} & $N_{\rm H}$ (10$^{22}$ cm$^{-2}$) & $7.34_{-0.41}^{+0.39}$ & $7.22_{-0.36}^{+0.35}$ & $7.2_{-0.3}^{+0.4}$ \\

		& $CvrFract$ & $0.957_{-0.024}^{+0.015}$  & $0.947_{-0.035}^{+0.022}$ & $0.946_{-0.015}^{+0.025}$ \\
		\hline
		
		\texttt{xstar} & $column$ ($10^{23}$)& $1.23_{-0.15}^{+0.15}$ & $1.28_{-0.16}^{+0.17}$ & $1.29_{-0.14}^{+0.17}$ \\ 
		
		& $rlog\xi$ & $4.11_{-0.09}^{+0.1}$ & $4.11_{-0.09}^{+0.13}$ & $4.12_{-0.08}^{+0.12}$\\
		\hline
		
		\texttt{Gaussian} & $LineE$ (keV)& $1.71_{-0.013}^{+0.013}$ & $1.71_{-0.013}^{+0.014}$ & $1.71_{-0.014}^{+0.013}$\\
		
		& $norm$ & $0.12_{-0.02}^{+0.02}$ & $0.11_{-0.021}^{+0.022}$ & $0.11_{-0.019}^{+0.026}$ \\
		\hline
		
		\texttt{relxillNS}& $q_1$ & $3^*$ & $5^*$ & $7^*$ \\

        & $q_2$ & $3^*$ & $5^*$ & $7^*$ \\
        
        & $kT_{bb}$ & $1.11_{-0.032}^{+0.126}$ & $1.11_{-0.051}^{+0.103}$ & $1.11_{-0.047}^{+0.088}$ \\
		
		& $log(\xi)$ & $2.48_{-0.17}^{+0.26}$ & $2.47_{-0.22}^{+0.16}$ & $2.52_{-0.18}^{+0.09}$ \\
		
		& $A_{Fe}$ (solar) & $0.97_{p}^{+1.09}$ & $0.72_{p}^{+1.61}$ & $0.59_{p}^{+1.44}$ \\
		
		& $log(n_{\rm e})$ (cm$^{-3}$) & $17.11_{p}^{+1.32}$ & $16.91_{p}^{+1.35}$ & $16.91_{p}^{+0.85}$ \\
		
		& $norm$ & $0.007_{-0.002}^{+0.002}$ & $0.009_{-0.003}^{+0.002}$ & $0.01_{-0.004}^{+0.002}$ \\
		\hline
		
		\texttt{simplcutx} & $\Gamma$ & $3.42_{-0.33}^{p}$ & $3.39_{-0.42}^{p}$ & $3.38_{-0.32}^{p}$ \\
		
		& $f_{\rm sc}$ & $0.067_{-0.024}^{+0.008}$ & $0.066_{-0.03}^{+0.013}$ & $0.067_{-0.023}^{+0.012}$ \\ 
		
		\hline
		
		\texttt{kerrbb} & $a_*$ & $0.928_{-0.022}^{+0.022}$ & $0.891_{-0.014}^{+0.013}$ & $0.878_{-0.005}^{+0.016}$\\
		
		& $i$ (degrees) & $56.95_{-1.75}^{+1.12}$ & $59.88_{-1.16}^{+0.42}$ & $61.24_{-0.98}^{+0.9}$ \\

        & $M_{\rm BH}$ ($\mathrm{M}_\odot$) & $10^*$ & $10^*$ & $10^*$ \\
		
		& $\dot{M}_{\rm BH}$ & $1.45_{-0.19}^{+0.16}$ & $1.7_{-0.11}^{+0.13}$ & $1.79_{-0.09}^{+0.15}$ \\

        & $d_{\rm BH}$ (kpc) & $10^*$ & $10^*$ & $10^*$ \\
		
		\hline
		
		\texttt{mbpo} & $d\Gamma_1$ & $0.209_{-0.024}^{+0.03}$ & $0.207_{-0.026}^{+0.031}$ & $0.208_{-0.026}^{+0.036}$ \\
		
		\textit{NICER} &$d\Gamma_2$ & $0.066_{-0.079}^{+0.057}$ & $0.066_{-0.076}^{+0.054}$ & $0.066_{-0.081}^{+0.052}$ \\
		
		&$E_{br}$ & $5.96_{-0.91}^{+0.76}$ & $5.98_{-0.86}^{+0.77}$ & $5.96_{-0.82}^{+0.77}$ \\
		
		& $N$ & $0.989_{-0.029}^{+0.015}$ & $0.989_{-0.026}^{+0.02}$ & $0.989_{-0.026}^{+0.02}$\\
		
		\hline
		\texttt{mbpo} \\
		\textit{NuSTAR} B & $N$ & $0.987_{-0.002}^{+0.002}$ & $0.987_{-0.002}^{+0.002}$ & $0.987_{-0.002}^{+0.002}$\\
		\hline

		& $\chi^2$/d.o.f & $938.2/908$ & $941/908$ & $940.1/908$  \\
		\hline\hline
	\end{tabular}
	\\
	\textit{Note.} Parameters with the symbol $p$ denotes parameter hits its upper or lower boundary.
\end{table*}

\end{document}